\documentclass[aps,prd,amsmath, twocolumn, superscriptaddress,nofootinbib,showpacs,longbibliography]{revtex4-1}
\usepackage{caption}
\usepackage[T1]{fontenc}
\usepackage[utf8]{inputenc}
\usepackage{lmodern}
\usepackage{mathtools}
\usepackage{enumitem}
\usepackage{verbatim}
\usepackage{multirow}
\usepackage[dvipsnames, usenames]{xcolor}
\definecolor{linkcolor}{rgb}{0.0,0.3,0.5}
\usepackage[hypertexnames=false, unicode, colorlinks=true, linkcolor=linkcolor,
citecolor=linkcolor, filecolor=linkcolor,urlcolor=linkcolor,
pdfusetitle]{hyperref}
\usepackage{orcidlink}
\usepackage[all]{hypcap}
\usepackage{graphicx}
\usepackage{amssymb}
\usepackage[normalem]{ulem} 
\usepackage{bm} 
\usepackage[english]{babel}
\usepackage{blindtext}
\usepackage{ragged2e}
\usepackage[labelfont={small,bf}, font=footnotesize, justification=justified,format=plain,singlelinecheck=off]{caption}

\usepackage{csquotes}
\usepackage{float}

\usepackage[font=footnotesize]{subcaption}
\usepackage{amsmath}
\usepackage{physics}
\usepackage{xspace}
\usepackage[]{mdframed}
\usepackage{orcidlink}
\usepackage{natbib}

\def\6{\partial}

\usepackage[letterpaper,top=2.5cm,bottom=2.5cm,left=1.5cm,right=1.5cm,marginparwidth=1.5cm]{geometry}

\begin{document}

\newcommand\myeq{\stackrel{\mathclap{\tiny\mbox{poles}}}{=}}



\title{Generalized Perturbed Kepler Problem: \\ Gravitational Wave Imprints from Eccentric Compact Binaries}

\author{Rajes Ghosh~\orcidlink{https://orcid.org/0000-0002-1264-938X}}
\email{rajes.ghosh@icts.res.in}
\affiliation{International Centre for Theoretical Sciences, Tata Institute of Fundamental Research, Bangalore 560089, India}

\author{R. Prasad~\orcidlink{https://orcid.org/0000-0002-6602-3913}}
\email{prasad.r@icts.res.in}
\affiliation{International Centre for Theoretical Sciences, Tata Institute of Fundamental Research, Bangalore 560089, India}

\author{Kabir Chakravarti}
\email{chakravarti@fzu.cz}
\affiliation{CEICO, FZU-Institute of Physics of the Czech Academy of Sciences,
Na Slovance 1999/2, 182 21 Prague 8, Czech Republic}

\author{Prayush Kumar~\orcidlink{0000-0001-5523-4603}}
\email{prayush@icts.res.in}
\affiliation{International Centre for Theoretical Sciences, Tata Institute of Fundamental Research, Bangalore 560089, India}

\date{\today}

\begin{abstract} 
Observations of astrophysical binaries may reveal departures from pure Keplerian orbits due to environmental influences, modifications to the underlying gravitational dynamics, or signatures of new physics. In this work, we develop a unified framework to systematically study such perturbations in the ambit of the perturbed Kepler problem and explore their impact on eccentric orbital dynamics and gravitational wave emission. Unlike traditional parametrized frameworks such as post-Newtonian and post-Einsteinian expansions, our approach offers a more source-specific modeling strategy, making it more natural to trace the physical origins of eccentric binary model parameters. Starting from a general perturbed potential, we derive the modified orbit and compute the associated gravitational fluxes and phase evolution, assessing their observational relevance for both current and future detectors. This framework thus offers a general and physically transparent toolkit for probing such subtle deviations from standard dynamics in gravitational wave data.
\end{abstract}

\maketitle
\begingroup
\renewcommand\thefootnote{}\footnotetext{\\First two authors contributed equally.}%
\addtocounter{footnote}{0}
\endgroup

\section{Introduction}
Gravitational wave (GW) astronomy has opened an unprecedented observational window into the strong-field regime of gravity, enabling direct tests of general relativity (GR)~\cite{LIGOScientific:2019fpa, LIGOScientific:2020tif, LIGOScientific:2021sio} and precise characterization of compact binary systems~\cite{LIGOScientific:2018mvr, LIGOScientific:2020ibl, KAGRA:2021vkt}. Since the landmark detection of \texttt{GW150914} by the LIGO-Virgo collaboration~\cite{LIGOScientific:2016aoc}, a growing catalog of binary mergers has started the era of multi-messenger astronomy~\cite{KAGRA:2013rdx}, provided novel insights into stellar population~\cite{LIGOScientific:2018jsj, LIGOScientific:2020kqk, KAGRA:2021duu}, and helped probe the dark universe~\cite{LIGOScientific:2017adf, LIGOScientific:2019zcs, LIGOScientific:2021aug}. With ongoing improvements in ground-based detectors and the advent of next-generation observatories such as Laser Interferometer Space Antenna (LISA)~\cite{LISA} and the Einstein Telescope (ET)~\cite{Abac:2025saz}, the next decade promises a dramatic increase in both the quantity and precision of detected signals, encompassing sources ranging from stellar binaries to extreme mass-ratio inspirals (EMRIs) and supermassive black hole mergers. In this high-precision era, accurate and flexible waveform models are essential for extracting astrophysical and fundamental physics information from GW signals.

The gravitational waveform emitted during a binary inspiral is determined by the underlying orbital dynamics. In the early to intermediate inspiral stages, the conservative motion is well approximated by the classical Kepler problem~\cite{Maggiore:2007ulw}, with relativistic corrections (e.g., precession, spin couplings) incorporated via post-Newtonian (PN) or self-force frameworks~\cite{Maggiore:2007ulw, Maggiore:2018sht, Blanchet:2013haa, Poisson:2011nh}. This conventional picture assumes two point masses in mutual Newtonian $1/r$ interaction. However, realistic astrophysical systems often involve perturbations to this idealized setup stemming from various sources, including the presence of nontrivial environments (e.g., dark matter, dense stellar regions, accretion disks)~\cite{Banerjee:2010, Rodriguez:2015oxa, Ishibashi:2020, Coogan:2021uqv, Rowan:2023, Sedda:2023big}, modifications to the underlying gravitational theory (e.g., higher-curvature terms and auxiliary field couplings)~\cite{Yunes:2009ke, Chamberlain:2017fjl, Liu:2018sia, Battista:2021rlh, DeFalco:2024ojf, Barsanti:2022ana, Payne:2024yhk, Trestini:2024zpi}, or potential new physics (e.g., deviation from Kerr paradigm and additional binary attributes like electric charge, magnetic dipoles, tidal effects etc)~\cite{Liu:2020bag, Carson:2020iik, Psaltis:2020ctj, Ioka:2000yb, Hinderer:2009ca, Bernaldez:2023xoh, magdip, Camilloni:2023rra, Henry:2023guc, Henry:2023len, Grilli:2024fds}.

In all these cases, binaries evolve under perturbed Keplerian potentials, and understanding their impact on orbital dynamics and the resulting GWs is the central objective of this work. As mentioned above, one major source of perturbation arises from the astrophysical environment, which affect the binary dynamics in two ways: (i) conservative modifications to the Newtonian potential due to the gravitational influence of the surrounding matter, and (ii) dissipative forces like dynamical friction or gas drag, which are typically non-central and velocity-dependent~\cite{Chandrasekhar:1943ys, Barausse:2014tra, Fedrow:2017dpk, Cardoso:2020lxx, Kocsis:2011dr, Yunes:2011ws, Toubiana:2020drf, Cardoso:2019rou, Cardoso:2022whc, Speeney:2022ryg, Boudon:2023vzl, Vijaykumar:2023tjg, Duque:2023seg}. While the former correction is often neglected for simplicity, it can nevertheless introduce deviations from the standard Keplerian ellipse and should ideally serve to define the new trajectory for computing the dissipative forces. For instance, binaries evolving in dark matter structure can experience long-term secular effects due to the motion in an effective metric (hence, an interaction potential) different from that in vacuum GR~\cite{Speeney:2022ryg, Cardoso:2021wlq, Speeney:2024mas}, leaving distinct signatures in the GW emission. Especially for EMRIs, even small environmental effects can accumulate over the $10^4 - 10^6$ orbits observed by LISA, leading to a measurable dephasing effect. Analytically modeling of such effects via perturbations to the Newtonian potential provides a clean and controlled approach to assess their influence.

Another motivation of considering perturbed Keplerian dynamics stems from potential modifications to GR and the Kerr paradigm, or in the presence of additional binary attributes like magnetic dipole and tidal fields~\cite{Yunes:2009ke, Chamberlain:2017fjl, Liu:2018sia, Barsanti:2022ana, Payne:2024yhk, Carson:2020iik, Psaltis:2020ctj, Ioka:2000yb, Hinderer:2009ca, Bernaldez:2023xoh}. Such features often manifest themselves by modifying the two-body interaction potential beyond the Newtonian term, providing a potential smoking gun for uncovering unknown physics beyond the standard GR picture~\cite{Peters:1963ux}. For example, in neutron star binaries, tidal fields contribute $\sim 1/r^6$ corrections~\cite{Hinderer:2009ca}, while compact objects with intrinsic magnetic dipole moments can introduce dipolar interactions of the form $\sim 1/r^3$~\cite{Ioka:2000yb, magdip}. In fact, one may consider other multipole-multipole interactions as well that can potentially affect inspiral dynamics. In particular, permanent $(l,l')$-pole coupling produces a long-range potential term $\sim 1/r^{l+l'+1}$ [e.g., permanent dipole ($l=1$)-dipole ($l'=1$) interaction producing a $1/r^{3}$ potential]. Whereas the induction of an $l'$-pole on a permanent $l$-pole produces a long-range potential term $\sim 1/r^{2(l+l'+1)}$ [e.g., a spherically symmetric mass ($l=0$) placed inside an external tidal field ($l'=2$) produces a $1/r^{6}$ potential]. By embedding such general corrections into a Unified Perturbed Keplerian (UPK) framework, we can study their influence and mutual interplay on binary dynamics without committing to a more involved and often-unknown relativistic formulation, while still retaining sufficient accuracy for early-to-intermediate inspiral phases.

Of particular importance is understanding how these perturbations affect key GW observables, most notably, the phasing of the waveform.
Since phase evolution encodes the cumulative dynamics of the binary and is the most accurately measured quantity, even small deviations from the ideal Keplerian motion can lead to detectable imprints. This sensitivity becomes especially pronounced in the context of space-based detectors such as LISA, which will monitor binaries over extended timescales spanning months to years~\cite{LISA}. With LISA, we expect to observe a diverse range of binaries, from pairs of white dwarfs and neutron stars to various black hole systems, including supermassive and intermediate-mass pairs, as well as extreme mass ratio inspirals (EMRIs). The long-duration signals from such sources make them powerful probes of subtle deviations in the gravitational potential. Another proposed space-based mission, Deci-hertz Interferometer Gravitational wave Observatory (DECIGO) is currently planned for launch in the $2030$s. We expect it to detect a large number of binaries in the frequency band between $0.1$ and $10$ Hz~\cite{Kawamura:2011zz}, further enhancing our observational reach. However, since LISA and DECIGO are not expected to be operational until late $2030$s~\cite{LISA2,Kawamura:2011zz}, it is crucial to assess the extent to which current ground-based detectors can already probe these effects. 

With these motivations, we propose a UPK framework to characterize the modified eccentric binary motion, compute the gravitational waveform in the quadrupole approximation\footnote{GW radiation depends on both conservative and dissipative dynamics. For the conservative part, we quantify deviations from the Keplerian case within the UPK framework, while for the dissipative part, we adopt a minimalist approach, retaining the GR quadrupole formula. Although generic modifications may induce extra scalar (say) radiation, this assumption remains well motivated, as argued in~\cite{Cardenas-Avendano:2019zxd}. Broadly, three cases exist: (i) dissipative corrections appear first at a lower PN order than conservative ones, (ii) they first arise at the same PN order, or (iii) conservative corrections appear first. If case (i) holds, such theories would already be strongly constrained by observations. Following Ref.~\cite{Cardenas-Avendano:2019zxd}, we therefore restrict to cases (ii) and (iii), where keeping the leading dissipative sector GR-like is consistent. \label{foo1}}, and estimate corrections to observable quantities such as the GW phase estimation. We present both analytic estimates and numerical computations to quantify the resulting phase shifts and analyze their detectability and highlight the potential for source characterization relevant for both present and future detectors. As a special case, we apply this formalism to compare with a few known cases, like binaries with magnetic dipoles~\cite{Ioka:2000yb}, and tidal interactions~\cite{Hinderer:2009ca, Bernaldez:2023xoh}. Several future prospects of this general framework are also outlined.

We must emphasize that the UPK framework developed in this paper provides a complementary alternative to traditional approaches, such as parametrized post-Newtonian (PPN)~\cite{Will, Mishra:2010tp} and parametrized post-Einsteinian (ppE) expansions~\cite{Carson:2020iik}. While those frameworks are invaluable for testing GR, they are often agnostic to the physical origin of deviations  (focusing primarily on quasi-circular inspiral) and typically introduce a number of abstract coefficients directly in the GW waveform to be constrained observationally. By contrast, our unified approach encodes deviations in an eccentric waveform that can be traced back to some physical origin, making our framework more pliant to modeling a wide range of astrophysical setups. Furthermore, the standard approach for solving the perturbed Kepler problem is via the method of ``osculating elements''~\cite{gravity}, which tracks the slow evolution of the orbital parameters under perturbations. This formalism is particularly powerful when the perturbations are small and act over many orbits. Nevertheless, we shall follow a more direct method used by several earlier studies, including that by Hinderer et al~\cite{Hinderer:2009ca}. It allows us to maintain a transparent connection between orbital dynamics and waveform observables. The novelties of our approach are, however, to
\begin{itemize}
    \item extend the analysis of Ref.~\cite{Hinderer:2009ca} to accommodate a unified perturbing potential $\delta V(r) \sim \sum_{k \geq 1} N_k/r^k$ (see Eq.~\eqref{pot} for more details) and obtain the deviated eccentric orbit $r(\phi)$,
    
    \item explicitly compute the associated gravitational wave fluxes for this general potential and connect them to central observable quantities like dephasing. This establishes a tractable bridge between source modeling and waveform generation, and paves the way for constructing fast, semi-analytic templates tailored to need-based phenomenological scenarios in future applications.
\end{itemize}

It is also important to note that our analysis is carried out within the Newtonian (slow-motion and weak-gravity) regime, focusing on deviations from the Keplerian potential in the early to mid inspiral stages. For relativistic systems, higher post-Newtonian (PN) corrections become significant due to the high orbital velocities involved. In such cases, the perturbative terms derived here has to be supplemented with waveform models that account for PN terms, such as \texttt{TaylorF2ecc}~\cite{Moore:2016qxz} or their time-domain variants like \texttt{TaylorT4ecc}~\cite{Huerta:2016rwp}, to study waveform deviations. Similarly, for late inspiral phase, strong-gravity effects become important as well and, depending on the mass-ratio, one has to supplement our framework with higher-order PN terms, self-force corrections~\cite{Nagar:2018zoe, vandeMeent:2023ols}, or effective-one-body waveforms calibrated to numerical relativity~\cite{Pompili:2023tna}. For weakly-gravitating non-relativistic sources ($v \ll c$), however, non-inclusion of such additional terms remains sufficiently robust to fully capture the relevant dynamics. This is the domain of applicability of our present work. We emphasize that the present work represents only the first effort to construct a unified parametrized framework for the perturbed Kepler problem. We plan to address various shortcomings of the present analysis in future works.

\section{Formulation of the problem}
Let us consider a binary system of two point masses $m_1$ and $m_2$ evolving under a central potential $V(r)$, which for the Newtonian case is given by $V_0(r)=-N_0\, \mu/r$, where $N_0=G\, m$ and $m=m_1+m_2$ is the total mass. It is well known that such a system can be reduced to an effective $1$-body planar motion for the reduced mass $\mu=m_1m_2/m$ in the center-of-mass reference frame $(r, \phi)$ with a Lagrangian, 
\begin{equation} \label{Lagrangian}
    \mathcal{L}=\frac{1}{2}\mu(\dot{r}^2+r^2\, \dot{\phi}^2)-V(r).
\end{equation} 
In the Keplerian case, the bound conservative dynamics of such a system is described by ellipses parameterized by the semi-major axis $a$ and eccentricity $e\in[0,1)$, or equivalently, by the total energy $E_0<0$ and angular momentum $h_0$ of the orbit.

However, as discussed in the Introduction, one acquires departures from this simple picture in the presence of perturbations stemming from environmental effects, beyond-GR signatures, or any new physics:
\begin{equation} \label{pot}
    V(r) = -\frac{N_0\, \mu}{r}+\sum_{k \geq 1} \frac{\epsilon\, N_k\, \mu}{r^k}.
\end{equation}
Here, $\epsilon$ is an order-counting parameter, and the coefficients $\{N_k\}$ ($k$ can take non-integer values as well) prescribe the deviation parameters, which are completely unspecified at this stage and can be fixed for a given astrophysical setup. We shall treat these additional parameters to be small with respect to the Newtonian contribution and neglect terms like $\mathcal{O}(N_k^2)$, and $\mathcal{O}(N_k\, N_l)$. So, the domain of validity of our calculation will be $|N_k| \ll N_0\, r^{k-1}$, i.e.,
\begin{equation} \label{validity}
    \begin{split}
        \frac{|n_k|}{\nu_k} \ll 1,\, \, \text{with}\, \, \nu_k &:= \left(\frac{m}{M_\odot} \right)\, \left(\frac{r}{r_\odot} \right)^{k-1} \\
        &= \left(\frac{f_\odot}{f}\right)^{2(k-1)/3}\left(\frac{m}{M_\odot}\right)^{(k+2)/3},
    \end{split}
\end{equation}
where $n_k=N_k/(c^2\, r_\odot^k)$ is a dimensionless quantity, $r_\odot=G\, M_\odot/c^2$, $f_\odot=64.3\, \text{kHz}$, and $f=\pi^{-1}\sqrt{N_0/r^3}$ representing the GW frequency at a binary separation $r$. Typically, the binary separations during the early and mid-inspiral are much larger than $r_\odot$. For simplicity, we have put a book-keeping parameter $\epsilon$ with $N_k$'s, so that the aforementioned approximation scheme reduces to neglecting $\mathcal{O}(\epsilon^2)$-terms. One must set $\epsilon=1$ in all final expressions.

Such perturbations to the Kepler problem as modeled in Eq.~\eqref{pot} by introducing corrections to the ``radial force'' is truly ubiquitous in astrophysical scenarios. For example, axially symmetric perturbations (like those arising from extended mass distributions, modifications to the Kerr paradigm, or new binary attributes) naturally contribute only radial forces, preserving the conservation of angular momentum and ensuring the motion remains planar. These conservative perturbations are derivable from a scalar potential (i.e., $\vec{F} = -\vec{\nabla} V$), which enables analytically tractable descriptions across a broad class of astrophysical systems that we investigate in this work. In such radially-perturbed settings, conservation of angular momentum (both in magnitude and direction) of the binary follows as an immediate consequence. We emphasize, however, that there exist physically relevant scenarios where the angular momentum of the binary alone may not be conserved, either due to dissipative losses or angular momentum transfer to ambient environment. Such mechanisms introduce additional dynamical features, including, for example, out-of-plane precession. A detailed treatment of these more general situations lies beyond the scope of the present analysis and will be explored in future works.

\begin{figure}[htbp!]
    \centering
    \includegraphics[width=\linewidth]{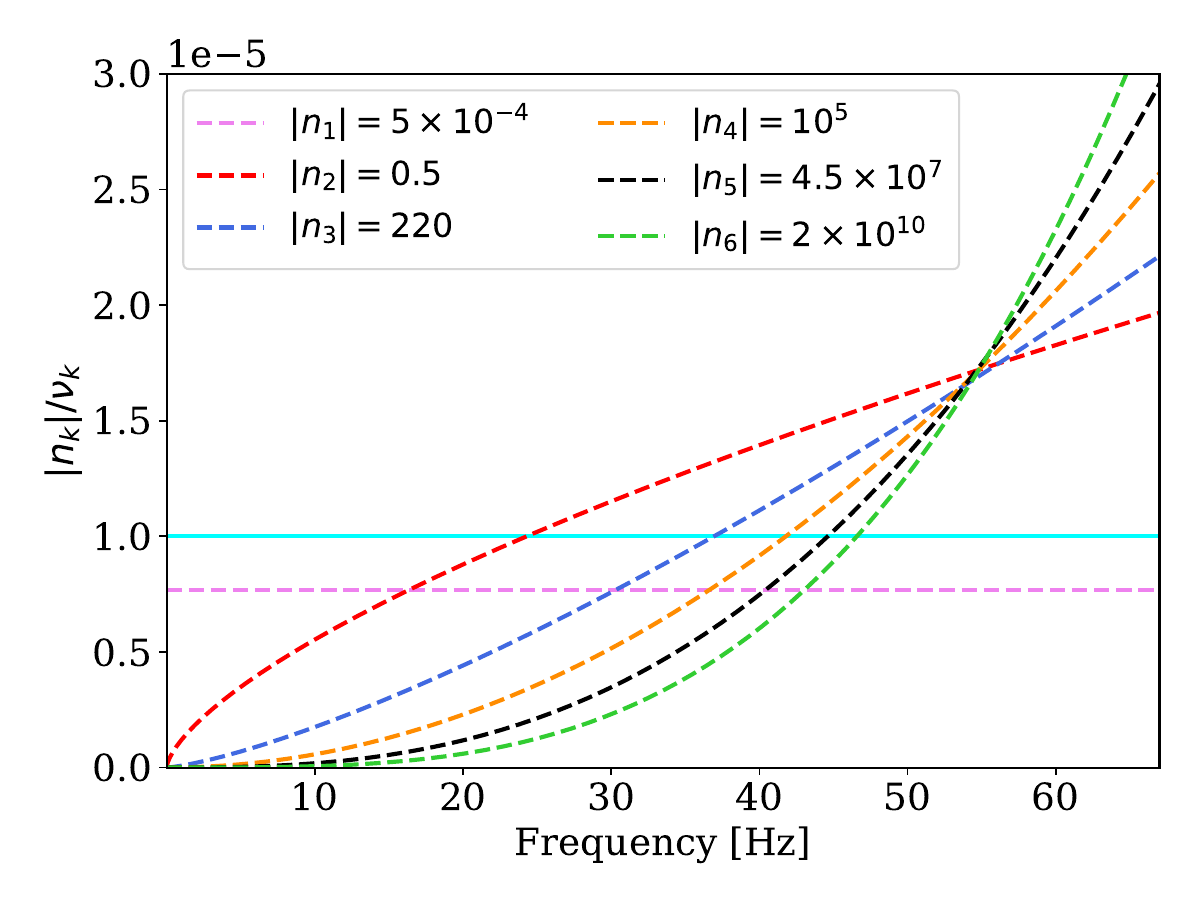}
    \caption{Variation of the ratio $|n_k|/\nu_k$ with frequency for $k=\{1,\cdots,6\}$ for some arbitrarily chosen values of $n_k$ (only one at a time) for a \texttt{GW150914}-like system with masses $36 M_\odot$ and $29 M_\odot$. The solid cyan line represents a reference line where the ratio is $10^{-5}$, which is well within the domain of our approach.}
    \label{fig:ratio}
\end{figure}

Fig.~\ref{fig:ratio} illustrates that the contribution from each $n_k$, quantified by the ratio $|n_k|/\nu_k$, which generally grows with frequency across all $k$, except for $k=1$ that remains constant\footnote{In Fig.~\ref{fig:ratio}, our choices of $|n_k|$ are phenomenologically motivated and and primarily intended for illustrative purposes. However, their values are not completely arbitrary and unphysical. As a concrete example, consider a compact neutron star binary with aligned magnetic fields of $B \sim 10^{15}\, \text{G}$ for each component. In our framework, this corresponds to $k=3$ with a value of $|n_3| \sim 500$ and $|n_k|/\nu_k = 10^{-3}$ at $f=10\, \text{Hz}$. This motivates our representative choice of $|n_3| = 220$. Also, since different $n_k$ originates from distinct astrophysical mechanism, they may not obey fixed scaling laws and in fact, a given system may only produce a certain set of $k$-perturbations.}. Importantly, these contributions stay within the regime of validity of our perturbative treatment. This suggests that ground-based detectors like LIGO may have potential to constrain some of these $n_k$ parameters depending on their relative strength with respect to the Keplerian background. However, the prospect will be particularly favorable for LISA, which will ``see'' the GW signal for a very long time and will be able to probe the slightest deviation from the Keplerian picture. Notably, the behavior of the $k=2$ curve deviates from the others, and several curves with $k > 1$ intersect each other at some high frequencies. These features are a direct consequence of the ratio $|n_k|/\nu_k$ scaling as $f^{2(k-1)/3}$.

Before proceeding further, let us also highlight two important points. First, while we focus on central potential perturbations, $\phi$-dependent interactions can also be accommodated by averaging them (will be denoted by $\bar{N_k}$) over the background Keplerian orbits ($\epsilon=0$), as argued in Refs.~\cite{Ioka:2000yb, magdip}. This orbital-averaged treatment is well justified for slowly evolving binaries during early and intermediate inspiral phases. Second, beyond environmental and modified gravity effects, specific $N_k$ terms may arise from intrinsic binary attributes as well, such as electromagnetic charges ($N_1$-term)~\cite{Benavides-Gallego:2022dpn, Chen:2022qvg}, magnetic dipoles ($N_3$-term)~\cite{Ioka:2000yb, Henry:2023guc, Henry:2023len}, or tidal interactions ($N_6$-term)~\cite{Hinderer:2009ca, Bernaldez:2023xoh}. 

\section{Conservative Dynamics}
In the presence of these perturbations, the binary motion is no longer described by Keplerian ellipses (except the $k=1$ case). In fact, the new orbital equation satisfies the following Euler-Lagrange equation:
\begin{equation} \label{orbiteq}
    \frac{d^2r}{d\phi^2}-\frac{2}{r}\left(\frac{dr}{d\phi}\right)^2-r+\frac{N_0\mu^2}{h^2}r^2=-\sum_{k\geq1}\frac{\epsilon\, k\, \bar{N_k}\, \mu^2}{h_0^2\, r^{k-3}},
\end{equation}
where we have used the angular momentum conservation equation $\mu\, r^2\, \dot{\phi}=h$ (with $h(\epsilon=0)=h_0$) to replace $\dot{\phi}$ in terms of $r$ and neglected $\mathcal{O}(\epsilon^2)$-terms. Also, as discussed earlier, we have replaced all deviation parameters in terms of the Keplerian orbital-averaged quantities $\bar{N}_k$. 

We can explicitly check that for the unperturbed case ($\epsilon=0$), the Keplerian ellipse $r_0(\phi):= r\left(\epsilon=0, \phi\right) = a(1-e^2)/(1+e\, \cos\phi)$ 
represents the bound orbit solution, where $a(1-e^2) = h_0^2/(N_0 \mu^2)$. However, in the presence of generic $\bar{N}_k$'s, the solution will be perturbed from an ellipse. Let us consider the following ansatz for the perturbed orbit parameterized in terms of the same $(a,e)$ of the unperturbed orbit and the multipoles $\{\bar{N}_k\}$:
\begin{equation} \label{perturbed}
    r(\phi)=\frac{h^2}{N_0\mu^2(1+e\cos\phi)} \left[1+\epsilon\, \delta r(\phi) \right]\, .
\end{equation}
Physically, $\delta r(\phi)$ captures the modifications of the original Keplerian orbit parametrized by $(a,e)$ as the perturbations are turned on. We also demand that $r(\phi)$ should be an even function, as it should be invariant under our choice of angular orientation, and that $\phi=0$ can be chosen arbitrarily. These conditions are enough for a unique determination of the orbit (up to an overall scale fixed by the orbital angular momentum $h$). With the above ansatz, Eq.~\eqref{orbiteq} boils down to 
\begin{equation} \label{perteq}
    \delta r''-\frac{2e \sin\phi}{1+e \cos\phi} \delta r'+\frac{\delta r}{1+e \cos\phi}=-\sum_{k\geq1} \frac{c_k}{(1+e \cos\phi)^{2-k}}, 
\end{equation}
where $c_k = k(\bar{n}_k/\nu_k) \equiv k \bar{N_k} \omega^{2(k-1)/3}/N_0^{(k+2)/3}$ with $\omega=\sqrt{N_0/a^3}$ as GW angular frequency. Note that the characteristic orbital speed is $v=(N_0 \omega)^{1/3}=(\pi N_0 f)^{1/3}$ (since orbital frequency is half of the GW frequency, $f=2f_{orb}$). Thus, the terms proportional to $c_k$ represent the $(k-1)$-th PN order. Now, using the ``variation of parameters'' method~\cite{Nagle:2012}, one can
calculate $\delta r(\phi)$ as\footnote{Apart from those generated by perihelion precession, all terms in $\delta r$ are invariant under a combined action of $\phi \to \pi+\phi$ and $e \to -e$. Precession is also the reason why a term linear in $e$ appears in $\delta r$. Such terms disappear for $k=1$, a case with no perihelion precession, as shown in Eq.~\eqref{Dp}. This is expected because for $k=1$, the potential is still proportional to $1/r$ and, hence, the orbit remains a closed ellipse, as written explicitly in Eq.~\eqref{ellipse1}.}
\begin{equation} \label{drq}
    \begin{split}
        &\delta r \approx -e c_1\cos\phi(1-e\cos\phi) - \sum_{k\geq1}c_k \left\{1+\frac{e}{2}\left[(k-3)\cos\phi \right. \right.\\
        &\left. \left. +(k-1)\phi \sin\phi\right] + \frac{e^2}{6}\left[2(k-1)(k-2)-(k^2-7)\cos^2\phi \right. \right.\\
        &\kern14em \left. \left. -\frac{3(k-1) \phi}{2} \sin2\phi\right]\right\},
   \end{split}
\end{equation}
up to quadratic order in $e$, thereby neglecting terms proportional to $\mathcal{O}(\bar{n}_k e^3)$ besides neglecting $\mathcal{O}(\bar{n}_k \bar{n}_l)$. We shall follow this approximation scheme throughout the paper for simplicity.

For more details on the derivation of $\delta r(\phi)$ see Appendix-\ref{appa}. Also, note that the quantity ($e$) appearing in the above equation is the eccentricity of the unperturbed orbit. In terms of the perturbed orbit, however, $e$ can be thought of as a parameter specified via the energy $E$ and the angular momentum $h$. In fact, since the new orbit is not a conic section, the standard notion of eccentricity does not hold. However, in Appendix-\ref{appb}, we have proposed a new measure of orbital deformation from a circle that works for any orbit $r(\phi)$ ``close'' to an ellipse.

\subsection{Some Properties of the perturbed orbit}
We now discuss a few interesting properties of the new perturbed orbit given by Eqs.~\eqref{perturbed} and \eqref{drq}. First note that, unlike the Kepler problem, the perturbed orbit is not closed but undergoes in-plane perihelion precession (except for $k=1$). In fact, we expect this in the absence of Laplace-Runge-Lenz vector due to deviation from Newtonian $1/r$ potential. To quantify the precession rate, let us compute after how much angular rotation $(2\pi+\epsilon\, \Delta \phi)$ a point $r(\phi)$ on the orbit comes back to its initial position. One obtains the following expression for the perihelion precession rate 
\begin{equation} \label{Dp}
    \Delta \phi \approx \pi \sum_{k\geq1} c_k\, (k-1)\, ,
\end{equation}

\begin{figure}
    \centering
    \includegraphics[width=\linewidth]{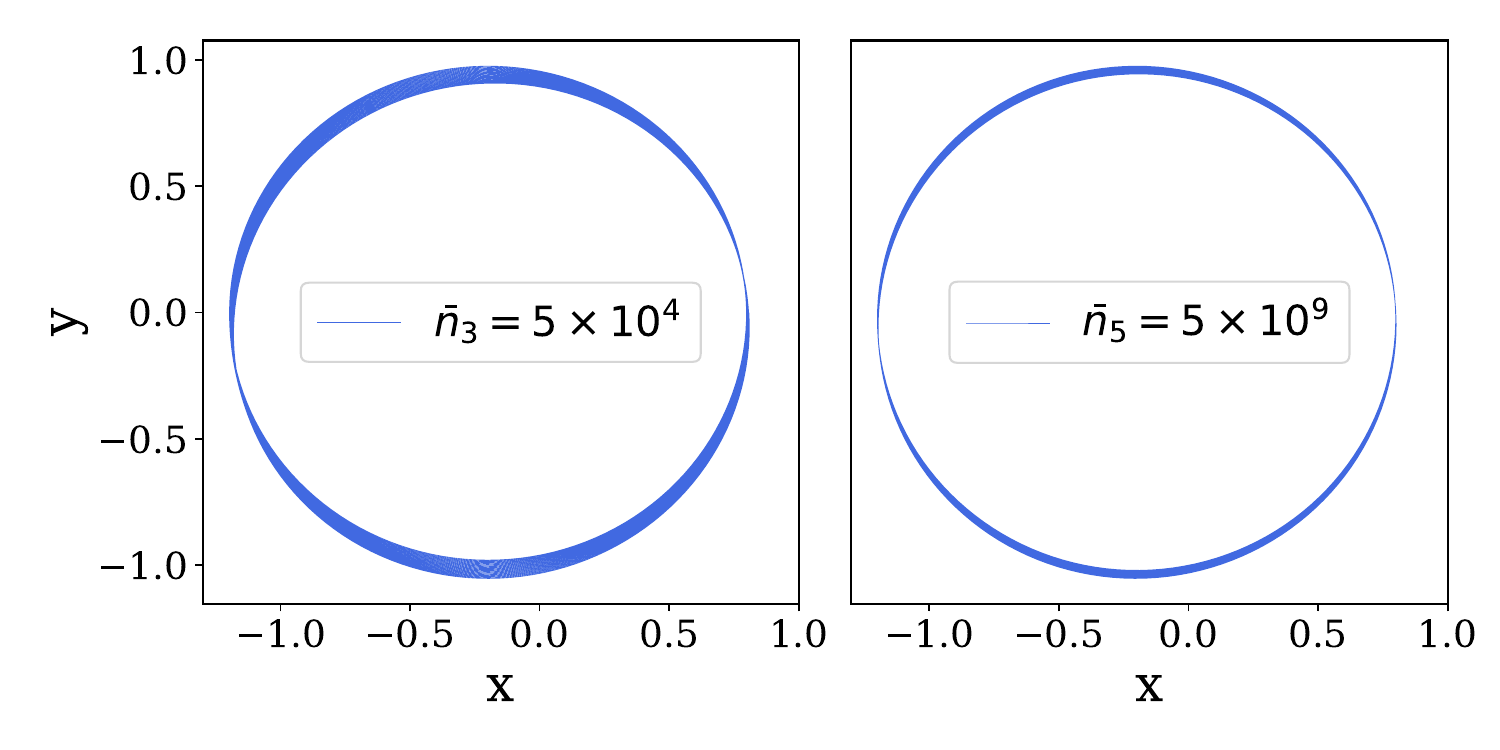}
    \caption{Orbits ($x=r(\phi) \cos\phi,\, y=r(\phi)\sin\phi$) with $r(\phi)$ given by Eq.~\eqref{neworbit} for $k=\{3,5\}$ (only one at a time) and $e=0.2$. Twenty cycles are shown and orbit sizes are normalized to unity ($\hat{a}=1$).}
    \label{fig:orbits}
\end{figure}

\noindent
which we have also checked directly from Eq.~\eqref{orbiteq} via Poincar\'e–Lindstedt method~\cite{Verhulst}. This motivates us to rewrite the orbit in the following suggestive manner so that the precession rate become very apparent:
\begin{equation} \label{neworbit}
    \begin{split}
        r(\phi) = \frac{\hat{a}(1-e^2)}{1+e\cos\left[\phi\left(1-\epsilon \frac{\Delta \phi}{2\pi} \right) \right]}\left[1+\epsilon\, \delta r_2(\phi) \right]\, ,
    \end{split}
\end{equation}
where $\hat{a}= a(h/h_0)^2$ and $\delta r_2$ has the same expression as $\delta r$ in Eq.~\eqref{drq} without the so-called secular terms proportional to $\phi$. Fig.~\ref{fig:orbits} shows the orbital precession for two cases with $k = \{3,5\}$ with an eccentricity of $e=0.2$. In this plot, we have chosen some large (unreasonably) values for $\bar{n}_k$ to make the orbital precession visually apparent. In actual astrophysical setup, their values will be much smaller and as a result, the corresponding precession will also be very tiny. One can also define a ``size'' of the orbit as
\begin{equation} \label{size}
    \begin{split}
        &\tilde{a}:=\frac{r(\phi=0)+r(\phi=\pi)}{2} \\
        &= \hat{a}\Bigg[1+2\epsilon e^2 c_1-\epsilon \sum_{k\geq1} c_k \left\{1+\frac{(k-4)(k-5)}{6}e^2 \right\} \Bigg]\, .
    \end{split}
\end{equation}
Here, we have used $\phi=\{0,\pi\}$ as the standard reference points on the orbit to calculate its size.

\subsection{Scheme for orbital averages}
For what follows, we need to set up a scheme for the orbital average $\langle Q \rangle$ of a quantity $Q(\phi)$ over a time period $T$. Since the  perturbed orbit precesses (except for $k=1$), its radial and angular period do not match. Hence, we choose to perform the orbital average over a radial period $T$ spent between $\phi=0$ to $\phi=2\pi+\epsilon \Delta \phi$:
\begin{equation} \label{avegrage}
    \begin{split}
        \langle Q \rangle &=\frac{1}{T}\, \int_{0}^{T} dt\, Q = \frac{1}{T}\, \int_{0}^{2\pi+\epsilon \Delta \phi} \frac{d\phi}{\dot{\phi}}\, Q \\
        &\approx \frac{1}{T}\, \int_{0}^{2\pi} \frac{d\phi}{\dot{\phi}}\, Q+\epsilon \left(\frac{\Delta \phi}{T}\right) \frac{Q(2\pi, \epsilon=0)}{\dot{\phi}(2\pi,\epsilon=0)}\, .
    \end{split}
\end{equation}
Note that, in the absence of perihelion precession,  radial and angular averages agree as expected. Now, in order to calculate $T$ and $\dot{\phi}$ in terms of relevant orbital quantities, we now proceed as follows. Integrating $\dot{\phi}$ over a time period $T$, we obtain 
\begin{equation} \label{hT}
    \begin{split}
        &T = \frac{h^3}{N_0^2 \mu^3} \int_{0}^{2\pi+\epsilon \Delta \phi}d\phi\, \frac{1+2\, \epsilon\, \delta r(\phi)}{(1+e\, \cos\phi)^2}, \\
        &= \frac{2\pi h^3}{N_0^2 \mu^3(1-e^2)^{3/2}}\left[ 1 + 3\epsilon\, e^2 c_1+2\, \epsilon\, \sum_{k\geq1} I_3(k)\right]\, ,
    \end{split}
\end{equation}
where we have used
\begin{equation} \label{I3}
    \begin{split}
        &I_3(k) := \frac{c_k}{4} \left\{(k-5)-e^2(k^2-6k+11)\right\}.
    \end{split}
\end{equation}
Also, we have $a(1-e^2) = h_0^2/(N_0 \mu^2)$, which implies
\begin{equation} \label{Th}
    \begin{split}
        &T = \frac{2\pi}{\hat{\omega}}\left[1+ 3\epsilon\, e^2 c_1+2\epsilon\, \sum_{k\geq1} I_3(k) \right],\\
        &\dot{\phi}=\frac{\hat{\omega} (1+e\, \cos\phi)^2}{(1-e^2)^{3/2}}\left[1-2\epsilon\, \delta r(\phi)\right],
    \end{split}
\end{equation}
where $\delta r(\phi)$ is given by Eq.~\eqref{drq}, $\hat{\omega}=\omega(h_0/h)^3=\sqrt{N_0/\hat{a}^{3}}$ with $\hat{a}= a(h/h_0)^2$. These scaling rules are somewhat expected as for a Keplerian orbit, we have $a \propto h^2$ and $\omega \propto a^{-3/2} \propto h^{-3}$. Note that we still have no relation connecting the angular momentum of the perturbed orbit ($h$) to its Keplerian counterpart ($h_0$), which leaves the overall scale of the orbit arbitrary. Since the orbital size directly affects the orbital dynamics and GW observables, one must adopt a calibration prescription. Different choices may be appropriate depending on the astrophysical system under consideration. In this work we restrict ourselves to two representative cases:\\

{\bf Case-a:} We set $h=h_0$, which holds if the perturbations do not change the angular momentum of the old Keplerian orbit. A particularly relevant astrophysical situation is when the perturbation builds up adiabatically compared to the orbital timescale. In such cases, the action variables of the system (and therefore the angular momentum) are approximately conserved, e.g., binaries embedded in a dark-matter spike whose density grows slowly around a supermassive black hole~\cite{Cipollina, Gondolo:1999ef, Sadeghian:2013laa}. In these scenarios, identifying $h$ with its Keplerian counterpart provides a natural calibration condition.\\

{\bf Case-b:} We consider the condition $\langle \dot{\phi} \rangle=\langle \dot{\phi} \rangle|_{\epsilon=0}$. This choice equates the orbital-averaged angular frequencies for the perturbed orbit with the old Keplerian one. This prescription is directly motivated from the observational perspective that GW detectors measure the phase evolution $\Phi(t) = 2\phi(t)$, relating the orbital-averaged GW frequency as $\omega = 2 \langle \dot{\phi} \rangle$. Thus, at a given observed GW frequency, it is natural to demand that the perturbed and Keplerian binaries share the same orbit-averaged angular frequency, so that any additional phase difference reflects the effect of the perturbation. In addition, this choice reproduces the same energy flux/phasing results as derived by in Refs.~\cite{Ioka:2000yb, Hinderer:2009ca} for the corresponding cases of circular binaries, which we shall discuss later in detail. This choice implies the following relation
\begin{equation} \label{calibrate}
    h=h_0 \left[1- \epsilon\, e^2 c_1-\frac{2\, \epsilon}{3}\, \sum_{k\geq1} I_3(k) +\frac{\epsilon\, \Delta \phi}{6\pi}\right]\, .
\end{equation}
We emphasize that other prescriptions are also possible, and different choices will in general lead to different predictions for observables such as the inspiral dephasing. The selection of the appropriate calibration condition must therefore be tailored to the physical nature of the perturbation under consideration. 

\section{Dissipative dynamics}
So far we have only considered the conservative features of the binary dynamics. However, such an orbiting system must radiate GWs, which will cause the binary to merge eventually. In order to compute the associated fluxes, we shall work under Einstein's quadrupolar approximation. Our goal will be to find the modifications in the standard Matthew-Peters formula~\cite{Peters:1963ux} due to the potential perturbations introduced in the previous section.

At this point, it is important to highlight a key observation. As we will see by incorporating dissipative fluxes in the following subsections, our framework becomes sufficiently complete to capture the effects of beyond-GR modifications or potential new physics (e.g., deviations from the Kerr paradigm and additional binary attributes). The situation, however, becomes more intricate when perturbations arise from an ambient environment, for which our current approach includes their effects only partially via the radial-potential corrections. This is because, in an environmental setting, both the conservative dynamics and the GW flux acquire additional contributions from mechanisms such as dynamical friction and accretion. Unlike the former cases, these effects may introduce velocity-dependent forces and extra dephasing contributions that depend sensitively on the environmental density profile, $\rho(r)$, as discussed in Ref.~\cite{Chandrasekhar:1943ys, Kavanagh:2020cfn, Nichols:2023ufs}. Incorporating such contributions in the UPK framework would require a general, parametrized form of $\rho(r)$, which introduces additional complexity and warrants a separate, detailed analysis. Therefore, in the present work, we focus exclusively on the influence of velocity-independent potential-based perturbations. Nonetheless, in Appendix-\ref{appc}, we outline a method to include other effects within our framework for any given $\rho(r)$.

\subsection{GW flux}
Taking into account the modified orbit given by Eqs.~\eqref{perturbed} and \eqref{drq}, the mass quadrupole tensor $M_{ij}$ takes the form
\begin{equation} \label{Mij}
    \begin{split}
        M_{ij} = &\frac{\mu \hat{a}^2 (1-e^2)^2}{(1+e\cos\phi)^2} \begin{pmatrix}
    \cos^2\phi & \sin\phi \cos\phi & 0\\
    \sin\phi \cos\phi & \sin^2\phi & 0\\
    0 & 0 & 0
    \end{pmatrix} \\
    &\kern11em \times \left[1+2\epsilon\, \delta r(\phi)\right]\, ,
    \end{split}
\end{equation} 
Now, following the discussion in Footnote-\ref{foo1}, we use the quadrupole formula for the average radiated power ($P_{GW}=dE_{GW}/dt$)~\cite{Maggiore:2007ulw},
\begin{equation} \label{P}
    \begin{split}
        &P_{GW}=\frac{G}{5c^5}\left\langle\dddot{M}_{11}^2+\dddot{M}_{22}^2+2 \dddot{M}_{12}^2-\frac{1}{3}\left(\dddot{M}_{11}+\dddot{M}_{22} \right)^2\right\rangle,
    \end{split}
\end{equation}
where the orbital averages are performed using Eq.~\eqref{avegrage}. One can similarly compute the angular momentum flux ($dJ_{GW}/dt$), whose explicit expression however we do not use in this work \footnote{For the sake of completeness, we quote only the final expression of the angular momentum flux as $\dot{J}_{GW}=\dot{J}_{GW}^{(0)}\left[1-\frac{5e^2}{4} \epsilon c_1 +\epsilon \sum_{k\geq1} c_k\left\{\frac{13-k}{2}+\frac{e^2(24k^2-73k+69)}{16} \right\} \right]$, with $\dot{J}_{GW}^{(0)}=\frac{32G^{7/2}m^{5/2}\mu^2}{5c^5 a^{7/2}(1-e^2)^2}\left(1+\frac{7e^2}{8}\right)$ being the corresponding Keplerian value.}. After some tedious algebra, one finally obtains:
\begin{equation} \label{power}
    \begin{split}
        &P_{GW} = \frac{32 G^4 \mu^2 m^3}{5 c^5 \hat{a}^5}f(e)\Bigg[1+\frac{37e^2}{12} \epsilon\, c_1\\
        &\kern4em +\epsilon \sum_{k\geq1} c_k \left\{8+\frac{e^2(24k^2-7k-54)}{12}\right\} \Bigg].
    \end{split}
\end{equation}
Here, $f(e) = (1-e^2)^{-7/2}(1+73e^2/24+37e^4/96)$ and we have kept terms up to $\mathcal{O}(e^2)$ for those multiplying $\epsilon$. At the end, we should set the book-keeping parameter $\epsilon$ to unity. Also, we have used the identity $G^p m^q/a^n=G^{p-n/3}m^{q-n/3}\omega^{2n/3}$. Note that we recover the standard Matthew-Peters formula~\cite{Peters:1963ux} when all $n_k$'s are set to zero.

\begin{figure*}
    \centering
    \includegraphics[width=0.9\textwidth]{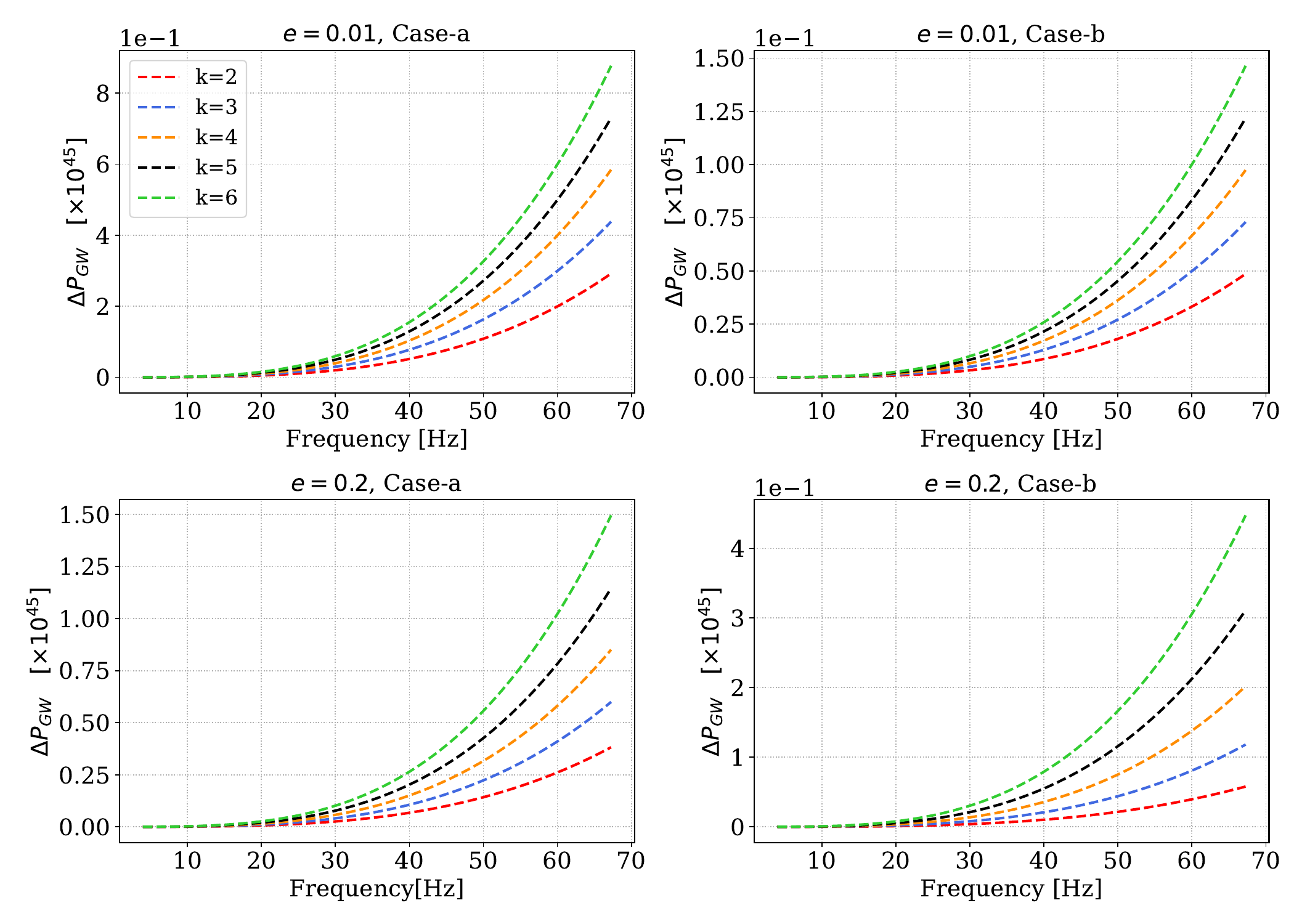}
    \caption{Variation of GW energy flux differences from the corresponding Keplerian value as a function of GW frequency for a \texttt{GW150914}-like system. We have chosen $\{k=2, \cdots,6\}$ (only one at a time) represented by the same colored lines as indicated in the upper-left subplot. The corresponding values of $\bar{n}_k$ are fixed such that $\bar{n}_k/\nu_k=10^{-5}$ at $f_{ISCO} = 67.26\, \text{Hz}$. } 
    \label{fig:powers}
\end{figure*}

Now, depending on the physical system at hand, we may be interested in (case-a) or (case-b) to fix $h$ in terms of $h_0$. For the first case ($h=h_0$), we can replace $\hat{a}$ by $a$ in the above equation to get the emitted power $P_{GW}^{(a)}$. However, for the other case, we need to use Eq.~\eqref{calibrate} to replace $\hat{a}$ with respect to $a=(N_0/\omega^2)^{1/3}$ and obtain  
\begin{equation} \label{P2}
    \begin{split}
        &P_{GW}^{(b)} = \frac{32 G^{7/3} \mu^2 m^{4/3}\omega^{10/3}}{5 c^5}f(e)\Bigg[1+\frac{157e^2}{12} \epsilon\, c_1 \\
        &\kern4em +\epsilon \sum_{k\geq1} c_k \left\{\frac{4}{3} +\frac{e^2(4k^2+113k-274)}{12}\right\} \Bigg].
    \end{split}
\end{equation}
Similarly, one can compute the angular momentum flux too, but we shall skip it for the purpose of the present paper. The variations of GW energy flux differences $\Delta P_{GW}=P_{GW}-P_{GW}(\epsilon=0)$ from the corresponding Keplerian value is showcased in Fig.~\ref{fig:powers} for a \texttt{GW150914}-like system with different choices of $\bar{n}_k$'s. Besides incorporating both cases-(a,b) defined earlier, we have also chosen two values of eccentricity $e=\{0.01,0.2\}$. The distinction between case-a and case-b is very clear, as is the variation caused by the change in eccentricity while remaining in our perturbative domain. Also, $\Delta P_{GW}$ in both cases takes positive values as the $k$-dependent pre-factors multiplying $c_k$ remains positive for the chosen parameters. 

Moreover, using energy balance ($\dot{E}_{orb}=-P_{GW}$) at the quadrupolar order, one can compute the rate of orbital shrinkage, where $E_{orb}$ is the orbital energy given by
\begin{equation} 
    \begin{split} \label{Eorb}
        E_{orb} &= \left[\frac{h^2}{2\mu r^2} -\frac{N_0 \mu}{r} - \sum_{k \geq 1}\frac{\epsilon\, \bar{N}_k\, \mu}{r^k} \right]_{\phi=0}\\
        &=-\frac{N_0 \mu}{2\hat{a}}\left[1-2\epsilon e^2c_1+2\epsilon \sum_{k\geq1} (1+k\, e^2) \frac{c_k}{k} \right].
    \end{split}
\end{equation}
As before, depending on the relation between $h$ and $h_0$ for case-a or b, the expression for $\hat{a}$ in terms of the parameters $(a,e)$ of the unperturbed Keplerian orbit will be different.
\begin{equation} 
    \begin{split} \label{Eorbb}
        E_{orb}^{(b)}=-\frac{N_0 \mu}{2 a}\Bigg[1-\frac{\epsilon}{3} \sum_{k\geq1}c_k \Bigg\{&\frac{2(2k-3)}{k} \\
        &+(k-1)(k-5) e^2\Bigg\} \Bigg].
    \end{split}
\end{equation}
Note that the change in the orbital energy can take both signs depending on the values of $k$. Also, from now on, we shall only focus on case-b alone for the purpose of the presentation. However, similar results can also be obtained for case-a.

\subsection{Electromagnetic flux}
Note that some terms in the perturbed potential can be partly contributed by electromagnetic multipoles of the binary components as well. For example, the $k=1$ ($k=3$) term in Eq.~\eqref{Lagrangian} can be partly sourced by electric monopole-monopole (magnetic dipole-dipole) interaction. In that case, such contributions will also give rise to electromagnetic radiation due to the accelerated motion of the binary. 

To incorporate their effects in the binary evolution, we may break $N_{1}\to N_{1}^O+N_{1}^E$ and $N_{3}\to N_{3}^O+N_{3}^E$ (or $c_1=c_1^E+c_1^O$ and $c_3=c_3^E+c_3^O$), where $E$ ($O$) denotes the electromagnetic (anything else) pieces. Since all forms of energy couple to gravity in the same way, both $O$ and $E$-pieces contribute to gravitational radiation. However, only the $E$-pieces will contribute to the electromagnetic radiation.
For illustration purpose, we shall assume that all other multipole interactions are negligible except the two cases discussed above, and consider the magnetic dipole moments $\vec{\mu}_{1,2}$ to be aligned and always perpendicular to the orbital plane. Although a generalization beyond this simple setup is straightforward, it is not too important for our study.

For electric monopole, $N_1^E=-k_e\, q_1 q_2/\mu$, or equivalently $c_1^E=-k_e\, q_1 q_2/(G \mu m)$ with $k_e=(4 \pi \epsilon_0 )^{-1}$ in SI ($k_e=1$ in CGS), and the net electric dipole moment of the binary is $\vec{p}=q_1 \vec{r}_1+q_2 \vec{r}_2$, where $q_{1,2}$ are the component charges and $\vec{r}_{1,2}$ are their instantaneous position vectors. We shall assume the charges to be small and neglect terms like $c_1^Ec_k$. In the center-of-mass frame it becomes $\vec{p}=q_e \vec{r}$ with $q_e=(q_1m_2-q_2m_1)/m$. Now, using the formula for electric dipole radiation~\cite{Benavides-Gallego:2022dpn}, we get
\begin{equation} \label{edipolerad}
    \begin{split}
        &\langle\dot{E}_{EM}^{k=1}\rangle= \frac{2 k_e}{3c^3} \langle\ddot{\vec{p}}^{\, 2}\rangle = \frac{2 k_e G^{2/3} m^{2/3}\omega^{8/3} q_e^2(1+3e^2)}{3c^3}.
    \end{split}
\end{equation}
Similarly, for the magnetic dipole radiation, we have $N_3^E=-k_m\, \mu_1 \mu_2/\mu$, or equivalently $c_3^E=-(3k_m \mu_1 \mu_2\omega^{4/3})/(\mu N_0^{5/3})$ with $k_m=\mu_0/(4\pi)$ in SI ($k_m=1$ in CGS). Then, the average power emitted is given by~\cite{Ioka:2000yb},
\begin{equation} \label{mdipolerad}
    \begin{split}
        &\langle\dot{E}_{EM}^{k=3}\rangle= \frac{4 k_m G^2 m^2\mu_e^2}{15c^5} \left\langle \frac{\dot{\vec{r}}^2-6\dot{r}(\hat{r}.\dot{\vec{r}})+9\dot{r}^2}{ r^6}\right\rangle \\
        &\kern3em = \frac{4 k_m G^{2/3} m^{2/3} \omega^{14/3} \mu_e^2 (1+15e^2)}{15c^5},
    \end{split}
\end{equation}

where $\mu_e=(\mu_1m_2-\mu_2m_1)/m$ and $\mu_{1,2}$ are component magnetic moments. Also, since $q_e^2$ and $\mu_e^2$ are $\mathcal{O}(\epsilon)$-quantities, we have neglected terms like $q_e^2 c_k$ and $\mu_e^2 c_k$. This is exactly the reason why the above formula (similarly Eq.~\eqref{edipolerad}) is independent of the potential correction introduced by our UPK framework.

\section{Dephasing}
In this section, we shall study the effects of perturbation in the GW phasing formula given by~\cite{Tichy:1999pv}
\begin{equation} \label{phasedd}
    \frac{d^2\psi}{df^2}=\frac{2\pi}{\dot{E}_{orb}}\frac{dE_{orb}}{df}=-\frac{2\pi}{\dot{E}_{GW}+\dot{E}_{EM}}\frac{dE_{orb}}{df}\, .
\end{equation}
While integrating this formula to obtain $\psi(f)$, we need to be mindful of eccentricity evolution~\cite{Peters:1964zz, Moore:2016qxz}, i.e., $e(f)\approx e_0\left(f_0/f\right)^{19/18}$, up to the leading order. Note that we can use this relation because the eccentricity appearing in all the previous formulas is that of the unperturbed Keplerian orbit. Moreover, for simplicity and the purpose of comparison with the standard result, we shall expand all quantities in eccentricity (till quadratic order) about their corresponding circular value, which is a bit different from what we were doing so far. We have the following Keplerian result

\begin{equation} \label{psi0}
    \psi_0=\psi_c\left(1-\frac{2355\,e_0^2}{1462\,\chi^{19/9}}\right),\,\, \psi_c=\frac{3\,c^5}{128\,G^{5/3}\,m^{2/3}\,\mu\, \omega^{5/3}},
\end{equation}
where $\chi=\omega/\omega_0$. However, for non-zero values of $\epsilon$, we have the phase $\psi=\psi_0+\epsilon\left(\psi_1+\psi_2\right)$:
\vspace{0.5cm}
\begin{widetext}
\begin{equation} \label{psi123}
    \begin{split}
        &\psi_1= -\psi_c\left[\frac{5 \left(k_e q_e^2 c^2+7 k_m \mu_e^2 \omega^2\right)}{84G^{5/3}m^{2/3}\mu^2\omega^{2/3}}+\frac{40}{3}\sum_{k\geq1|k\neq5}\frac{(2k-1)c_k}{(k-5)(2k-7)}-\frac{80c_5}{3}\left(1-\log \chi\right)\sum_{k\geq1}\delta_{k5}\right],\\
        &\psi_2=-\frac{5\, e_0^2\, \psi_c}{ \chi^{19/9}}\left[-\frac{\left(41261\, k_e q_e^2 c^2-9016\, k_m \mu_e^2 \omega^2\right)}{1069376\, G^{5/3}m^{2/3}\mu^2\omega^{2/3}} +\frac{471}{731} c_1+\sum_{k\geq1}\frac{(12k^3-98k^2+313k-855)}{(3k-20)(6k-49)}c_k\right],
    \end{split}
\end{equation}
\end{widetext} 
where $c_k = [k\, c^2\, r_\odot^k\, \bar{n}_k\, \omega^{2(k-1)/3}]/N_0^{(k+2)/3}$ as mentioned before, $\delta_{k5}$ is the Kronecker delta function that is non-zero and unity only when $k=5$. Therefore, we have the following dephasing formula
$\delta\psi(f)=\epsilon\left(\psi_1+\psi_2\right).$ Then, the cumulative dephasing in the frequency range $[\omega_0,\omega_{f}]$ is given by
$\delta\Phi=\int_{\omega_0}^{\omega_f}\omega\, \delta\psi''(\omega)\, d\omega$~\cite{Tichy:1999pv}. The upper limit of this integration ($\omega_f$) should be chosen depending on the underlying detector band. However, for the purpose of presentation, we shall fix it to be the angular frequency corresponding to the innermost stable circular orbit (ISCO), i.e., $\omega_f=\omega_{ISCO}$. Using previous formulas, this can be calculated as, $\delta\Phi=\epsilon\left(\Phi_1+\Phi_2\right)|_{\omega_0}^{\omega_{ISCO}}$, where
\begin{widetext}
\onecolumngrid
\begin{equation} \label{phi123}
    \begin{split}
        &\Phi_1= \frac{5\psi_c}{126}\left[\frac{5 k_e q_e^2 c^2+14 k_m \mu_e^2\omega^2}{G^{5/3}m^{2/3}\mu^2\omega^{2/3}}-224\sum_{k\geq1}\frac{(2k-1)c_k}{2k-7}\right],\\
        &\Phi_2=\frac{5\, e_0^2\, \psi_c}{9 \chi^{19/9}}\left[-\frac{\left(1331 k_e q_e^2 c^2-184 k_m \mu_e^2\omega^2\right)}{704\, G^{5/3}m^{2/3}\mu^2\omega^{2/3}}+\frac{471}{17} c_1-\sum_{k\geq1}\frac{(12k^3-98k^2+313k-855)}{(3k-20)}c_k\right],
    \end{split}
\end{equation}
\twocolumngrid
\end{widetext}
Eqs.~\eqref{psi123}
and \eqref{phi123} represent the dephasing effects caused by generic potential-perturbations given by Eq.~\eqref{pot} and constitute two central equations in this work. Since dephasing is a directly measurable quantity, these results will be particularly important for constraining such perturbations from Keplerian dynamics using GW observations. We shall discuss various observational prospects in detail in a subsequent section. 

Before we move on, a few points are worth mentioning. Since $c_k \sim v^{2(k-1)}$, our perturbed Keplerian approach generates both even and odd order PN terms (as leading effects) in the dephasing for integer ($\mathbb{Z}$) values of $k$. One can obtain the other PN corrections (e.g., half-integer like in PPN framework) as well by considering $r^{-k}$ corrections to the Newtonian potential with fractional values of $k$ (e.g., $k \in \mathbb{Z}/2$). Another important consequence of the above formulas is that the $k$-dependent pre-factors multiplying $c_k$ can change sign depending on the choice of $k$ (for a fixed $\bar{N}_k$), thereby producing associated dephasing terms of opposite signs.  In fact, depending on the value of $k$, the variation net dephasing as a function of $f$ could be strikingly different, as shown in subsequent figures. This provides a tool to distinguish among various potential corrections if the GW signal is long enough to span over a reasonable range including low to high frequencies.

\section{Comparison with known results}
In this section, we shall compare the orbital energy, gravitational flux, and dephasing formulas with two known sub-cases:\\
\\
{\bf (i) Binary with magnetic dipoles $(G=c=k_m=1)$~\cite{Ioka:2000yb, Henry:2023guc}:} Using Eq.~\eqref{Lagrangian}, we have $\bar{N}_3=\mathcal{F}(\mu_1,\mu_2)/\mu$ for $k=3$, where $\mathcal{F}(\mu_1,\mu_2)=-\mu_1\mu_2$, $\mu = \eta m$. Here, $\mu_{1,2}$ represents the magnetic moments of the binary components. Then, using $\dot{E}_{orb}^{k=3}=-\dot{E}_{GW}^{k=3}-\dot{E}_{EM}^{k=3}$, Eq.~\eqref{power}, Eq.~\eqref{Eorbb}, Eq.~\eqref{mdipolerad}, and Eq.~\eqref{psi123} imply that (expanding around the circular result)
\begin{equation} \label{Ioka}
    \begin{split}
        &\dot{E}_{orb}^{k=3}=-\frac{32m^{10/3} \eta^2 \omega^{10/3}}{5} \Bigg[1+4\gamma_3+\frac{157e^2}{24}+\frac{617}{12}\gamma_3 e^2 \\
        &\kern14.5em +\frac{\mu_e^2(1+15e^2)}{24m^2 a^2 \eta}\Bigg], \\
        &E_{orb}^{k=3}=-\frac{\eta m^{5/3} \omega^{2/3}}{2}(1-2\gamma_3+4e^2\gamma_3), \\
        &\delta \psi^{k=3}=-100\gamma_3\left[1-\frac{711e_0^2}{3410\chi^{19/9}}\right]\psi_c\\
        &\kern7.5em-\frac{5 \mu_e^2\omega^{4/3}}{12m^{8/3}\eta^2}\left(1+
        \frac{69e_0^2}{682\chi^{19/9}}\right)\psi_c\,,
    \end{split}
\end{equation}
where $\gamma_3=\mathcal{F}(\mu_1,\mu_2)/(m^2a^2\eta)=\mathcal{F}(\mu_1,\mu_2)\omega^{4/3}/(\eta m^{8/3})$. These formulas match with the results of Refs.~\cite{Ioka:2000yb, Henry:2023guc} for circular ($e=0$) binaries.\\

{\bf (ii) Binary with tidal effects ($G=c=1$)~\cite{Hinderer:2009ca, Bernaldez:2023xoh}:} Comparing with Ref.~\cite{ Bernaldez:2023xoh} for $k=6$, we have the identification $\bar{N}_6=3m_2^2\lambda_1/(2\mu)$, or equivalently $c_6 = (9m_2^2\lambda_1 \omega^{10/3})/(\mu m^{8/3})$. Here, $\lambda_1$ corresponds to tidal deformability of the primary in the binary system. Then, using $\dot{E}_{orb}^{k=6}=-\dot{E}_{GW}^{k=6}$, Eq.~\eqref{power}, Eq.~\eqref{Eorbb}, and Eq.~\eqref{psi123} imply that (expanding around the circular result)
\begin{equation} \label{Tanja}
    \begin{split}
        &\dot{E}_{orb}^{k=6}=-\frac{32}{5}\mu^2 m^{4/3} \omega^{10/3}\Bigg[1+4\gamma_6+\frac{157e^2}{24}+\frac{979\gamma_6}{6}e^2\Bigg],\\
        &E_{orb}^{k=6}=-\frac{1}{2}\mu m^{2/3} \omega^{2/3}(1-3\gamma_6-5e^2 \gamma_6),\\
        &\delta \psi^{k=6}=-88\gamma_6\left[1+\frac{1305e_0^2}{2288\chi^{19/9}}\right]\psi_c\,,
    \end{split}
\end{equation}

\begin{figure*}
    \centering
    \includegraphics[width=\linewidth]{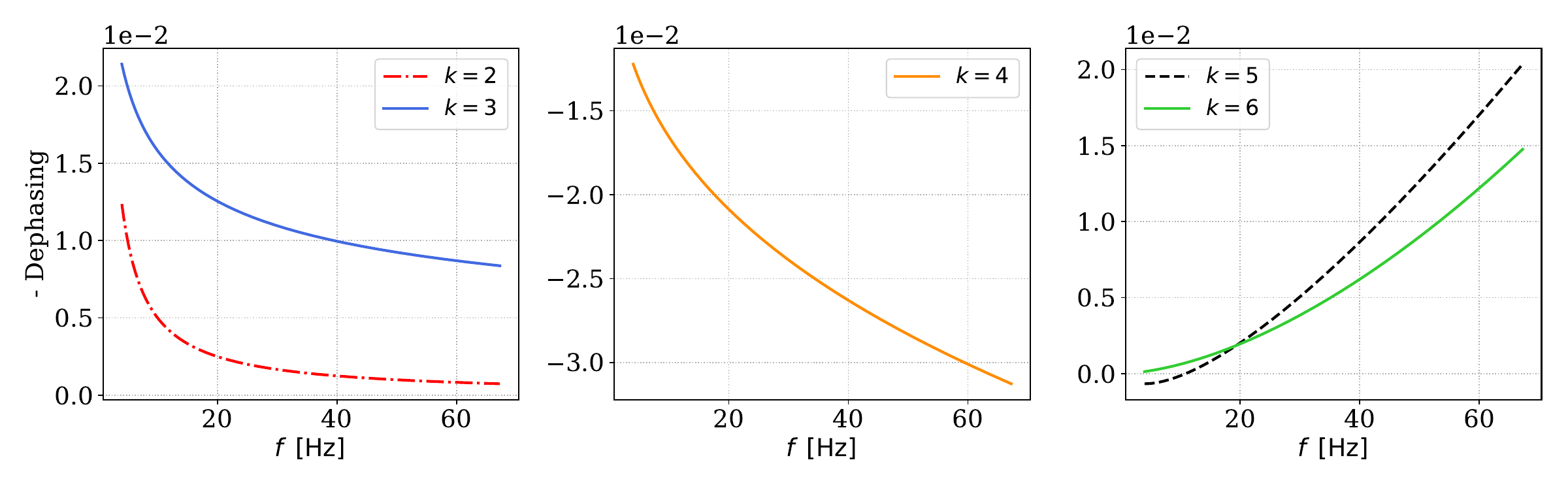}
    \caption{Instantaneous dephasing for a \texttt{GW150914}-like system for different $k$ values with $\bar{n}_k/\nu_k=10^{-5}$ at $f_{ISCO}$. The vertical axis shown is minus dephasing. We have chosen $e_0=0.1$ at $f_0=4$ Hz. For $k=2,3,5,6$, dephasing is negative, whereas for $k=4$ it is positive. For $k=2,3, 4$ it decreases with frequency, whereas for $k=5, 6$ it increases with increase in frequency. There is a switch over at $k=4$.}
    \label{fig:dephasing}
\end{figure*}

\begin{figure*}
    \centering
    \includegraphics[width=\linewidth]{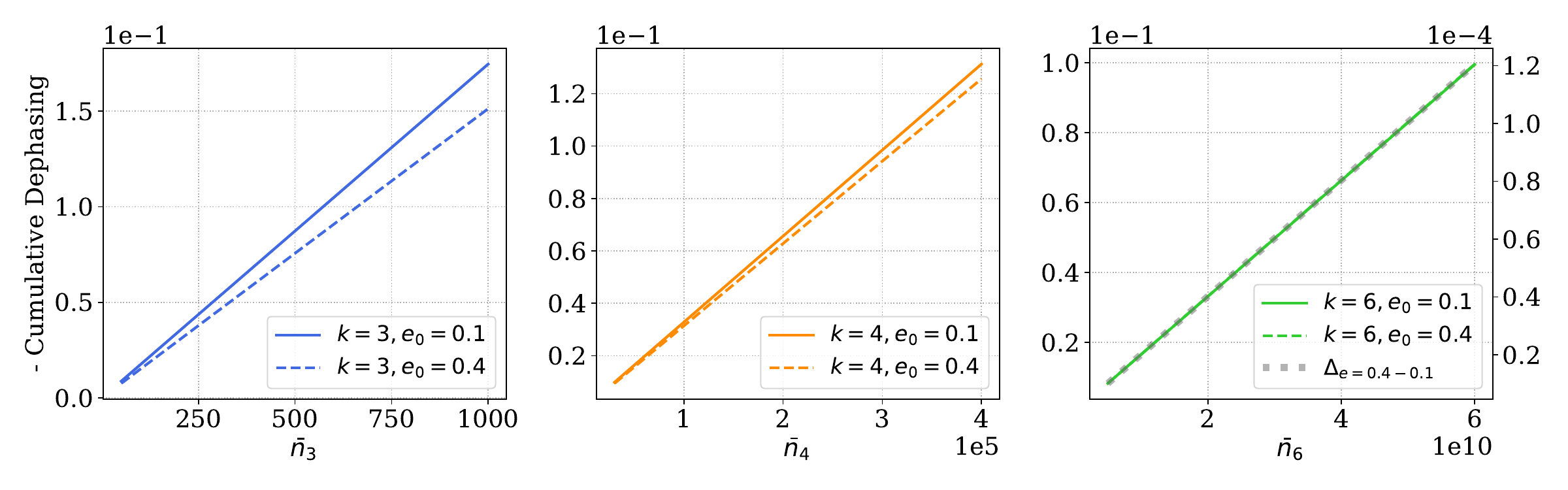}
    \caption{Cumulative dephasing for a \texttt{GW150914}-like system for different $k$ values as a function of different $\bar{n}_k$'s (only one at a time) and with $e_0=0.1$ (solid) and $e_0=0.4$ (dashed) at $f_0=4$ Hz. These $\bar{n}_k$-ranges are chosen in a window around the corresponding values of $\bar{n}_k$ for which $\bar{n}_k/\nu_k=10^{-5}$. For the cases shown, increasing eccentricity results in a decrease in the net cumulative dephasing. However, the case with $k=6$ shows a different nature, where eccentricity contribution enhances cumulative dephasing. Since variation for the chosen eccentricities is small, we depict the difference curve $\Delta_{e=0.4-0.1}$ as a gray dotted line (scale given on the RHS $y$-axis), which exhibits a positive value ($\sim 10^{-4}$) and a positive slope.}
    \label{fig:cumdephasingn}
\end{figure*}

where $\gamma_6 = (3m_2^2\lambda_1 \omega^{10/3})/(\mu m^{8/3})$. These formulas match the results presented in Refs.~\cite{Hinderer:2009ca, Bernaldez:2023xoh} for circular ($e=0$) binaries\footnote{For the purpose of illustration, we have set the induced quadrupole moment part and the tidal deformability $\lambda_2$ of the secondary component to zero. However, one can easily include their effects following Ref.~\cite{Hinderer:2009ca, Bernaldez:2023xoh}. The complete effect will be the sum of an induced term and a piece coming from the tidal effect on the secondary (by relabeling $1 \leftrightarrow 2$).}. However, the eccentric corrections of these expressions show deviations with respect to those in Ref.~\cite{Bernaldez:2023xoh} because of the following subtle reason. 
In our derivation of the orbit, we solved both $r$ and $\phi$ components of the Euler-Lagrange equations consistently and then use them to calculate fluxes. However, in Ref.~\cite{Bernaldez:2023xoh}, the authors derived the new radial equation by assuming the unperturbed Keplerian expression for $\dot{\phi}$ to hold true even for the perturbed orbits and as a result the angular momentum $\mu r^2(\phi) \dot{\phi}$ on the new orbit does not remain conserved, unlike our approach, even at the conservative level (i.e., with no GW emission).

\section{Observational Prospects}

\textit{(i) Phasing comparisons:} We present the instantaneous dephasing as a function of GW frequency ($f$) in Fig.~\ref{fig:dephasing} for a \texttt{GW150914}-like system with component masses $36 M_\odot$ and $29 M_\odot$, evaluated for different values of $k$, where $c_k$ corresponds to the $\bar{n}_k/\nu_k=10^{-5}$ at $f_{ISCO}=c^3/(6^{3/2}\pi G m)$ (all $\bar{n}_k$'s are chosen to be positive and $q_e=\mu_e=0$). The sign of the phase shift induced by the effect operating at a given $k$ is determined by the coefficient $(2k - 1)/[(k - 5)(2k - 7)]$ in  Eq.~\eqref{psi123}, corresponding to circular part (the expression of $\psi_1$), which dictates whether the contribution is positive or negative. For eccentric orbits, the eccentricity-dependent terms introduce additional offsets to this behavior. If we only consider integer values of $k$, $k = 4$ alone contributes a positive phase shift, while all other values contribute negative instantaneous shifts. Note that the value of instantaneous dephasing being $\mathcal{O}(10^{-2})$ is an artifact of our chosen $\bar{n}_k$ values. This scale can be adjusted by varying the strength of the perturbation.

The form of $c_k$ also determines whether the phase shift increases or decreases with frequency. This behavior is governed by the $\omega^{-5/3}$ dependence present in $\psi_{c}$, and is further modulated by the ratio of the power of $\omega$ in $c_k$ to that in $\psi_c$, which dictates the net scaling of the phase contribution. In fact, it is easy to observe (as also showcased in Fig.~\ref{fig:dephasing} with a few illustrative cases) that for $k \leq 4$, the instantaneous negative dephasing decreases with frequency. Whereas for $k > 4$, there is a turnover and instantaneous negative dephasing starts to increase with frequency. This fact is very crucial for observational purpose for the following reasons. First, detectors like LISA that are sensitive to low frequency regions will be particularly suited for probing potential-perturbations with $k \leq 4$, among which $k=4$ case is particularly distinct for generating a positive dephasing. In contrast, ground-based detectors like LIGO will be more sensitive to $k>4$ terms. Secondly, if there are two or more of such potential-deviations appearing together with multiple $k$'s some less and greater than four, an ambiguity in their detection can arise because of their mixed dependence with frequency.  

\begin{figure*}
    \centering
    \includegraphics[width=\linewidth]{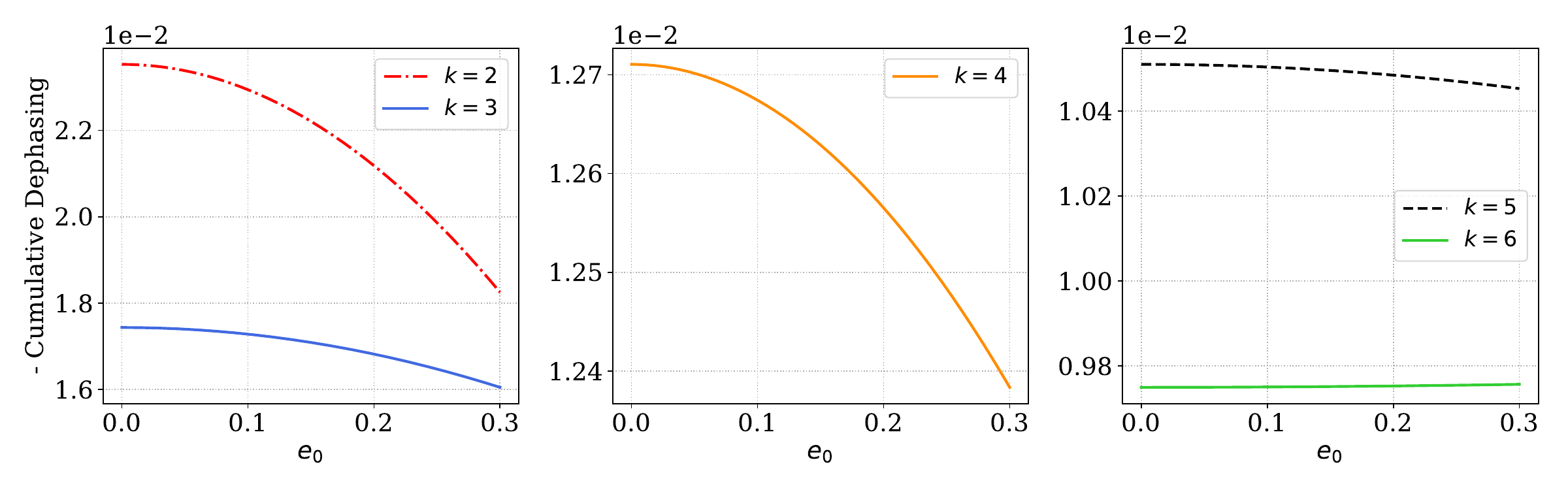}
    \caption{Cumulative dephasing for a \texttt{GW150914}-like system for different $k$ values as a function of initial eccentricity $e_0$ at $f_0=4$ Hz and $\bar{n}_k/\nu_k=10^{-5}$ at $f_{ISCO}$. The amount of absolute cumulative dephasing being $\mathcal{O}(10^{-2})$ is an artifact of the ratio $\bar{n}_k/\nu_k$ being $10^{-5}$.}
    \label{fig:cumdephasinge}
\end{figure*}

Fig.~\ref{fig:cumdephasingn} plots the negative cumulative dephasing (accumulated over about $400$ orbits starting from $4$ Hz till $f_{ISCO}$) for a range of $\bar{n}_k$ with two values of initial eccentricity $e_0=\{0.1,0.4\}$ at $f_0=4$ Hz. For the purpose of illustration, we have chosen $k=\{3,4,6\}$ cases alone and only one at a time. We observe that an increase of $e_0$ (remaining within our perturbative regime) decreases the amount of absolute cumulative dephasing for all cases considered except $k=6$. A potential correction with larger $k \in [1,6]$ diminishes this effect, which is evident when comparing the solid and dashed lines among the three panels in Fig.~\ref{fig:cumdephasingn}. Since $c_k$'s are proportional to $\bar{n}_k$'s, the linear nature of the plots are expected. Also, the monotonically increasing nature is understandable as the absolute value of cumulative dephasing is just the area under the instantaneous dephasing plots weighted by the time spent by the binary at that frequency. One can also notice that for the chosen ranges of $\bar{n}_k$'s in Fig.~\ref{fig:cumdephasingn}, the cumulative dephasing are reaching as high as $0.1$ radian (or beyond), an effect detectable (dephasing $\sim (\text{signal-to-noise ratio})^{-1}$) even in LIGO and will be particularly well measured in LISA. Fig.~\ref{fig:cumdephasinge} showcases the variation of negative cumulative dephasing with respect to initial eccentricity $e_0$ fixed at $f_0=4$ Hz. All plots (except for $k=6$, where the last term in the RHS of Eq.~\eqref{phi123} changes sign) show a decreasing nature with increasing values of $e_0$, suggesting that the binary is gathering less (more) absolute cumulative dephasing.

\begin{figure*}
    \centering
    \includegraphics[width=0.9\linewidth]{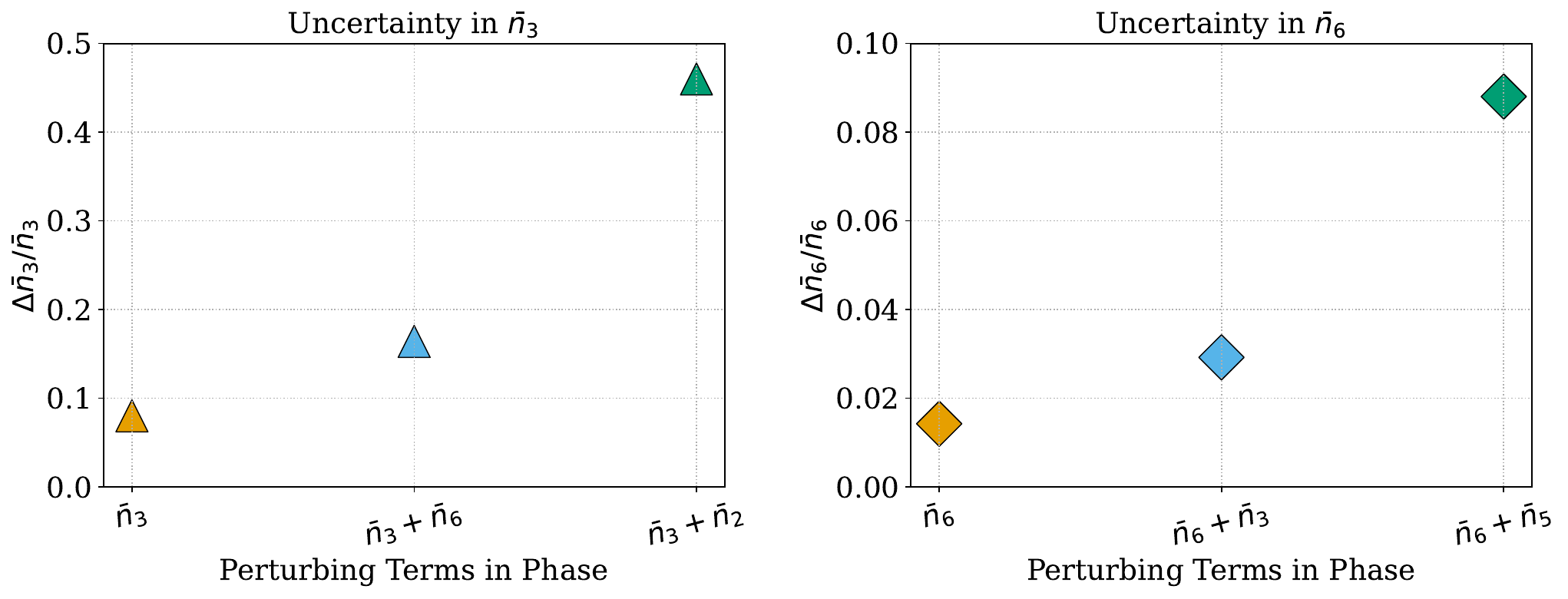}
 \caption{
 \textbf{(Left)} Fractional uncertainty in $\bar{n}_{3}$ estimated using Fisher analysis for \texttt{GW150914}-like system for three scenarios: (i) GW phase including only the $\bar{n}_{3}$ perturbation term, (ii) including both $\bar{n}_{3}$ and a $k < 4$ term, i.e., $\bar{n}_{2}$, and  (iii) including $\bar{n}_{3}$ along with a $k > 4$ term, i.e., $\bar{n}_{6}$. \textbf{(Right)} Fractional uncertainty in $n_{6}$ for the same three scenarios. The uncertainty is more strongly influenced when perturbation terms are exclusively from either $ k < 4 $ or $k > 4$, compared to when terms from both $k < 4$ and $k > 4$ are present together. This behavior arises since terms with similar frequency trends exhibit stronger mutual influence.}
    \label{fig:fisher_uncertainity}
\end{figure*}

\textit{(ii) Fisher analysis:} We now perform a Fisher analysis~\cite{Vallisneri:2007ev} to examine the interplay of perturbative terms with $k < 4$, $ k > 4$, and both together. The analysis is conducted using the \texttt{GWFish} package~\cite{Dupletsa:2022scg}. Our primary focus is on how these terms influence the parameter estimation uncertainty of a specific term.  Instantaneous dephasing reveals that perturbative terms in the two regimes exhibit distinct frequency-dependent behaviors. For this study, we consider a \texttt{GW150914}-like system, adopting component masses (36 and 29 $M_{\odot}$) and luminosity distance ($425 \, \mathrm{Mpc}$) as that event, along with fiducial values for sky location, geocentric time, polarization, and phase. In addition, we assign an eccentricity of $e_0=0.1$ at $4 \, \mathrm{Hz}$. We adopt a combined Einstein Telescope (ET)~\cite{Branchesi:2023mws, Abac:2025saz} and Cosmic Explorer (CE)~\cite{Evans:2021gyd, Evans:2023euw} detector network, which results in an SNR of $\sim 1858$. We estimate the parameter $\bar{n}_{3}$ by introducing perturbative terms from the set $\bar{n}_{k} \in \{\bar{n}_3, \, (\bar{n}_3, \, \bar{n}_6), (\bar{n}_3, \bar{n}_2) \}$. We consider perturbative potential magnitudes such that $\bar{n}_{k}/\nu_{k} = 10^{-5}$ at the ISCO. The fractional uncertainty obtained is illustrated in Fig.~\ref{fig:fisher_uncertainity}. When only $\bar{n}_3$ is present and estimated, the errors are the smallest. Inclusion of $\bar{n}_6$ alongside $\bar{n}_3$ increases the errors, while the combination of $\bar{n}_3$ with $\bar{n}_2$, a perturbative correction exhibiting a frequency trend similar to $\bar{n}_3$, results in substantially large errors.

Next, we consider $\bar{n}_{6}$, a term in the regime $ k > 4$, and estimate its measurement uncertainty by introducing perturbative terms from the set $n_{k} \in \{\bar{n}_6, \, (\bar{n}_6, \, \bar{n}_3), (\bar{n}_6, \bar{n}_5) \}$ such that $\bar{n}_{k}/\nu_{k} = 10^{-5}$ at the ISCO.  We find that the presence of $\bar{n}_5$, a perturbation that exhibits a similar increasing trend with frequency, affects the estimation of $\bar{n}_6$, leading to significant error. Estimating $\bar{n}_6$ when it is the sole perturbative term, the error is minimized, followed by the case with $\bar{n}_6$ and $\bar{n}_3$. It is important to note that we have fixed the magnitudes of terms to have comparable effects on the system's potential and emitted gravitational waves (see Figs.~\ref{fig:ratio} and \ref{fig:powers}). This step up highlights how different terms influence each other's estimation based on their frequency evolution trend. In real scenarios, both the magnitude of the perturbing potential and its frequency behavior would influence the estimation errors. Consequently, our ability to constrain the perturbative potentials in real events will be shaped by these subtleties.

An extensive analysis of these rich phenomenologies and their detectability/measurability requires a dedicated study. We aim to perform it in a future work with a few astrophysically motivated models, where deviations from Keplerian dynamics arise more naturally. Such analysis will help identify regions in parameter space where the perturbations are distinguishable from standard inspiral signals, and where degeneracies might arise, especially when multiple $r^{-k}$-perturbations are present. It will also clarify how the signal duration, signal-to-noise ratio, and detector configuration affect our ability to constrain specific types of perturbations.

\section{Discussions and Conclusions}
In this work, we developed a general framework (UPK) to study the GW signatures of binaries evolving under a perturbed Keplerian potential, arising from additional binary attributes, beyond-GR modifications, or other new physics such as environmental effects. By parameterizing deviations through physically motivated potential corrections, we derived their impact on the orbital motion and GW observables such as energy flux, with a particular emphasis on waveform phasing, an especially sensitive quantity in GW data analysis. Unlike conventional parametrized approaches such as PPN and pPE, our method maintains a direct connection between the model parameters and their physical origin, enabling a more interpretable and source-specific analysis. Furthermore, the framework captures the coupling between eccentricity and perturbations, extending its applicability to eccentric binary systems.

We observe distinct frequency trends for perturbative potentials with $k \leq 4$ and $k > 4$, with a flip around $k=4$. Furthermore, we find that eccentricity corrections generally reduce the impact of perturbative corrections compared to circular binaries, except for the case of $k \geq 6$, where this effect is reversed (i.e., increasing with $e_0$). Our findings highlight that even small deviations from the Keplerian motion can leave measurable imprints on the GW signal, which will become particularly pronounced in long-lived inspirals. The perturbed Kepler framework thus provides a versatile and computationally efficient tool for modeling a broad range of physically motivated scenarios, with potential applications in testing GR, probing astrophysical environments, and searching for exotic new physics using GW observations. 

It will be particularly interesting to consider loud LIGO events like \texttt{GW150914} and perform a rigorous Bayesian analysis to constrain the model parameters $\{\bar{n}_k\}$ with higher-order PN terms properly included in the waveform. As discussed earlier, the behavior of $k < 4$ dephasing terms is strikingly different from that with $k > 4$. This observation will be particularly important for distinguishing between various deviations from the Keplerian picture. Moreover, in this paper, we have only focused on the potential-corrections and their observational effects. For environmental effects, however, this picture will be incomplete if we do not include contributions from the dynamical friction or viscous drag too in the physical observables, especially in dephasing as outlined in Appendix-\ref{appc}. We leave these analyses for a future work. 

Other promising future directions include extending this approach to spinning binaries and incorporating non-central (broken angular momentum conservation) or time-dependent (broken energy conservation) deviations via the post-Keplerian parametrization~\cite{Damour:1991rd}. Both of these effects can arise in realistic physical systems due to friction and tilted binary motion with respect to a disk-like environmental structure, i.e., the binary will ``feel'' the environmental effects only periodically when the orbit crosses the environment.

\section{Acknowledgement}
We thank Ajith Parameswaran, Paolo Pani, and Sayak Datta for providing valuable comments. We extend our gratitude to the Astrophysical Relativity group at ICTS, for many engaging discussions and useful feedback. The work of K.C was partially supported by the PPLZ grant (Project No. 10005320/0501) of the Czech Academy of Sciences. We acknowledge support of the Department of Atomic Energy (DAE), Government of India, under project no. RTI4001. P.K. also acknowledges support by the Ashok and Gita Vaish Early Career Faculty Fellowship at the International Centre for Theoretical Sciences. This work makes use of \texttt{GWFish} package~\cite{Dupletsa:2022scg}, \texttt{Mathematica}~\cite{mathematica}, \texttt{numpy}~\cite{numpy}, \texttt{matplotlib}~\cite{matplotlib}, and \texttt{scipy}~\cite{scipy}.

\appendix
\section{Solving the orbit equation} \label{appa}
The homogeneous part of Eq.~\eqref{perteq} has two linearly independent solution $\delta r_{1,2}$ with Wronskian $W$:
\begin{equation} \label{homo}
    \delta r_1 = \frac{\cos\phi}{1+e\, \cos\phi},\, \delta r_2 = \frac{\sin\phi}{1+e\, \cos\phi},\, W = \frac{1}{(1+e\, \cos\phi)^2}\, .
\end{equation}
Then, using the variation of parameters method for solving a linear inhomogeneous differential equation, we obtain the following particular solution for Eq.~\eqref{perteq},
\begin{equation} \label{part}
    \begin{split}
        &\delta r_p = -\delta r_1 \int{d\phi\, \frac{\delta r_2}{W}\, S} + \delta r_2 \int{d\phi\, \frac{\delta r_1}{W}\, S}, \\
        &\implies \delta r_p = \sum_{k\geq1}c_k \left\{I_2(k,\phi)\, \delta r_1 - I_1(k,\phi)\, \delta r_2 \right\}\, ,
    \end{split}
\end{equation}
where we have used the shorthand $S$ to represent the source term in the RHS of Eq.~\eqref{perteq} and 
\begin{equation} \label{I12}
    \begin{split}
        &I_1(k, \phi) = \int{d\phi\, \cos\phi\, (1+e\, \cos\phi)^{k-1}}\, ,\\
        &I_2(k, \phi) = \int{d\phi\, \sin\phi\, (1+e\, \cos\phi)^{k-1}}\, .
    \end{split}
\end{equation}
Although the second integral can be performed exactly, the exact result for the first one is quite cumbersome! Hence, we shall only quote results up to the quadratic order in eccentricity:
\begin{equation} \label{I12e}
    \begin{split}
        &I_1 \approx \sin\phi+\frac{e(k-1)}{4}(2\phi+\sin2\phi)+\frac{e^2(k-1)(k-2)}{24}\times\\
        &\kern16em (9\sin\phi+\sin3\phi)\, ,\\
        &I_2\approx -\cos\phi-\frac{e(k-1)}{2}\cos^2\phi-\frac{e^2(k-1)(k-2)}{6}\cos^3\phi\, .
    \end{split}
\end{equation}
Note that $I_1$ is an odd function of $\phi$, whereas $I_2$ is an even function of $\phi$. Now, the complete solution for Eq.~\eqref{orbiteq} can be written as
\begin{equation}
    \begin{split}
        r = \frac{\hat{a}(1-e^2)}{1+e\, \cos\phi} &\Big[1 +\frac{\epsilon}{1+e\, \cos\phi}\, \Big(A \cos\phi+B\sin\phi\\
        &+\sum_{k\geq1}c_k \left\{I_2(k,\phi) \cos\phi - I_1(k,\phi) \sin\phi \right\}\Big)\Big] ,
   \end{split}
\end{equation}
where $\hat{a}=a(h/h_0)^2$, and $(A,B)$ are arbitrary constants multiplying the homogeneous solutions. Evenness of the solution demands, $B=0$. Also, if $k=1$, then the term in RHS of Eq.~\eqref{perteq} can be absorbed inside the $r^2$-term on LHS and one still gets an elliptical solution as
\begin{equation} \label{ellipse1}
    r_1(\phi) = \frac{\hat{a}(k=1)\, (1-e^2)}{1+e\, \cos\phi}(1-\epsilon\, c_1).
\end{equation}
We then obtain $A=-e\, c_1$. This fixes $A$ uniquely, as it is $k$-independent. Therefore, the final solution takes the form
\begin{equation} \label{comsol}
    \begin{split}
        &r(\phi) = \frac{\hat{a}(1-e^2)}{1+e\, \cos\phi}\Big[1 +\frac{\epsilon}{1+e\, \cos\phi}\, \Big(-e\, c_1 \cos\phi \\
        &\kern4em +\sum_{k\geq1}c_k \left\{I_2(k,\phi) \cos\phi - I_1(k,\phi) \sin\phi \right\}\Big)\Big] ,
   \end{split}
\end{equation}
Now, we can compare this solution with the ansatz Eq.~\eqref{perturbed} and obtain $\delta r(\phi)$ as given by Eq.~\eqref{drq} till the quadratic order in $e$.

\section{An analog of eccentricity $e$ for perturbed orbits} \label{appb}
Since the new orbit $r(\phi)$ given by Eq.~\eqref{perturbed} or equivalently in Eq.~\eqref{neworbit} is not an ellipse (in fact, not even a conic section), it does not have any eccentricity in the usual sense. However, since $r(\phi)$ deviates only perturbatively from the Keplerian ellipse parametrized by $(a,e)$, we should be able to define an analogous quantity (say ellipticity $\tilde{e}$) that quantify the deformation of the orbit from a circle. Our construction should be such that $\tilde{e} \to e$ if $\epsilon \to 0$.

Note that we can rewrite the Keplerian ellipse in terms of the Fourier modes as
\begin{equation} \label{kepel}
    r_0(\phi) = \frac{a(1-e^2)}{1+e\, \cos\phi}= \sum_{n\geq0} A_n(a,e)\, \cos(n \phi)\, .
\end{equation}
If we can construct a function $F(x)$ using the Fourier coefficients alone such that $F(A_n)=e$, then it can be used as the definition of ellipticity $\tilde{e}$ for the perturbed orbit. However, since a general orbit may also include perihelion precession, we should first go to a co-precessing frame of reference $\phi \to \Psi$ and then apply this construction on $r(\Psi)$.

With some trial and error, we have the following guess 
\begin{equation} \label{defe}
    \tilde{e} \equiv F(A_n) = \sqrt{\sum_{n\geq0}(-1)^{n+1}\, n\, \left( \frac{A_n}{A_0} \right)^2}\, .
\end{equation}
This guess is motivated by the following chain of reasoning. First, note that the correct limit $\tilde{e} \to e$ in the Keplerian case ($\epsilon=0$) dictates that both $\tilde{e}$ and $-\tilde{e}$ should represent the same ellipse (when $\epsilon=0$) connected by $\phi \to \phi+\pi$. Thus, a suitable choice which might be expressible in terms of the Fourier coefficients $A_n$ is $\tilde{e}^2$. Then, from Eq. \eqref{kepel}, it is also obvious that all $A_n$'s are proportional to the size of the orbit $a$. Thus, to separate out the eccentricity information, the construction should be done in terms of the ratio $(A_n/A_0)$, which is scale free. Moreover, since $\tilde{e}$ is a physical quantity, it should be invariant under the shift $\phi \to \phi+\phi_0$. Although under such a shift, Eq. \eqref{kepel} generates terms like $A_n'\, \cos(n \phi)$ and $B_n'\, \sin(n \phi)$, but the combination $A_n^2 = A_n'^2+B_n'^2$ remains invariant. This, in turn, motivates our choice $\tilde{e}^2=\sum_{n \geq 0} c_n\, (A_n/A_0)^2$, with some coefficients $c_n$ found to be equal to $(-1)^{n+1}\, n$.

Let us first check whether this construction works for the Keplerian ellipse. From Eq.~\eqref{kepel}, one finds that
\begin{equation} \label{ankep}
    A_0^{(0)}=a\, \sqrt{1-e^2},\, \, A_{n>0}^{(0)}=\frac{2\, (-1)^n\, e^n}{\left(1+\sqrt{1-e^2}\right)^n}\, A_0^{(0)}\, .
\end{equation}
Then, one can explicitly check that $\tilde{e}^{(0)} \equiv F(A_n^{(0)})=e$. Thus, our construction gives the correct result for the Keplerian ellipse. To the best of our knowledge, this construction of the deformation parameter has not been previously discussed in literature.

Now, we want to apply it for the perturbed orbit given by Eq.~\eqref{neworbit}. We can go to the co-precessing frame by the mapping $\phi \to \Psi[1+\epsilon\, \Delta \phi/(2\pi)]$ and the new orbit equation becomes $r(\Psi)$. Then, the Fourier coefficients can be calculated using the formulas
\begin{equation}
    \begin{split}
        &A_0=\frac{1}{2\pi}\int_0^{2\pi} r(\Psi)\, d\Psi\, ,\\
        &A_n=\frac{1}{\pi}\int_0^{2\pi} r(\Psi)\, \cos(n\Psi)\, d\Psi\, .
    \end{split}
\end{equation}
Moreover, under our approximation scheme, we shall also neglect terms proportional to $\mathcal{O}(\bar{n}_k \bar{n}_l)$, $\mathcal{O}(\bar{n}_k e^3)$ etc. Then, it is not hard to compute that
\begin{equation} \label{newe}
    \tilde{e} \approx e\left[1+\epsilon\, c_1+\sum_{k\geq1}\frac{\epsilon\, (k-3)}{2}\, c_k\right]\, .
\end{equation}

As a special case, when $k=1$ and all other coefficients vanish, one finds $\tilde{e}(k=1,c_{k \neq 1}=0) \approx e$, as expected. Surprisingly, the same holds for $k=3$ as well, $\tilde{e}(k=3,c_{k \neq 3}=0) \approx e$.

\section{Incorporating dynamical friction} \label{appc}
This Appendix outlines the method to include the effect of environmental dynamical friction (DF) in our framework. For simplicity, we illustrate it for an EMRI-type system. In this case, the smaller mass in the binary moves through environment with a density profile $\rho(r)$. As a result, it gets decelerated in the direction
of motion and loses both its kinetic energy and angular momentum.  Following Ref.~\cite{Chandrasekhar:1943ys}, the force due to DF is given by
\begin{equation} \label{dfforce}
    F_{DF}=\frac{4\pi G^2 \mu^2\, \mathrm{ln}\Lambda}{v^2}\, \rho(r),
\end{equation}
where  of the environment, $\mathrm{ln}\Lambda$ is the so-called Coulomb logarithm parameter, and $v$ is the velocity of the smaller mass. Now, since $\rho(r)$ itself $\mathcal{O}(\epsilon)$-quantity, the above formula coincides with that of Keplerian case till the first order. Hence, the velocity term can be replaced by its Keplerian value, i.e.,
\begin{equation} \label{v}
    v = \omega a\, \sqrt{\frac{1+2e \cos\phi+e^2}{1-e^2}}.
\end{equation}
Then, the averaged energy loss rate, due to DF is given by $\dot{E}_{DF}= \langle v F_{DF} \rangle$~\cite{Eda:2014kra}. This can be simplified, till the linear order in $\epsilon$, as
\begin{equation} \label{eddf}
    \dot{E}_{DF} \approx \frac{2 G^2 \mu^2 (1-e^2)^2 \, \mathrm{ln}\Lambda}{\omega a} \int_{0}^{2\pi}d\phi\, \frac{\rho(r)(1+e \cos\phi)^{-2}}{\sqrt{1+2e \cos\phi+e^2}}.
\end{equation}
Now, for a given model of $\rho(r)$, the above integral can be evaluated and then incorporated in the dephasing formula given by Eq.~\eqref{phasedd}.

\end{document}